\documentclass[]{IEEEtran}
\usepackage{cite}

\ifCLASSINFOpdf
   \usepackage[pdftex]{graphicx}
   \graphicspath{{images/}}
\else
\fi
\usepackage{amsmath}
\usepackage{amsfonts}
\usepackage{amssymb}
\usepackage{caption}
\usepackage{subcaption}
\usepackage{algorithm}
\usepackage{algorithmic}
\usepackage{color}

\usepackage{array}

\newcommand{\caliA}{\mathcal{A}}
\newcommand{\pCs}{\mathbf{Pr}(C\mid s)}
\newcommand{\pcs}{\mathbf{Pr}(c\mid s)}

\newcommand{\pCS}{\mathbf{Pr}(C\mid S)}
\newcommand{\newtext}[1]{{\color{black} #1}} 
\renewcommand{\epsilon}{\varepsilon}
\newcommand{\U}{\mathcal{U}}



\usepackage{color}
\usepackage{hyperref}
\usepackage[dvipsnames,svgnames,x11names,hyperref, enumerate]{xcolor}
\definecolor{blueish}{HTML}{007F99}
\definecolor{purpleish}{HTML}{72177A}
\hypersetup{
	colorlinks,
	citecolor=blueish,
	filecolor=red,
	linkcolor=purpleish,
	urlcolor=Blue
}

\graphicspath{{./images/}}

\DeclareMathOperator*{\argmin}{arg\,min}

\begin{document}
%
\title{Trace Reconstruction Problems in \\ Computational Biology}
%
%
%

\author{Vinnu~Bhardwaj$^{1}$,
        Pavel~A.~Pevzner$^{2*}$,
        Cyrus~Rashtchian$^{2, 3}$,
        and~Yana~Safonova$^{2}$
\thanks{$^{1}$Electrical and Computer Engineering Department, University of California San Diego, La Jolla, USA.}
\thanks{$^{2}$Computer Science and Engineering Department, University of California San Diego, La Jolla, USA.}
\thanks{$^{3}$Qualcomm Institute, University of California San Diego, La Jolla, USA.}
\thanks{$^{*}$Corresponding author: ppevzner@ucsd.edu}
}
\maketitle

\begin{abstract}
The problem of reconstructing a string from its error-prone copies, \textit{the trace reconstruction problem}, was introduced by Vladimir Levenshtein two decades ago. While there has been considerable theoretical work on trace reconstruction, practical solutions have only recently started to emerge in the context of two rapidly developing research areas: immunogenomics and DNA data storage. In immunogenomics, traces correspond to mutated copies of genes, with mutations generated naturally by the adaptive immune system. In DNA data storage, traces correspond to noisy copies of DNA molecules that encode digital data, with errors being artifacts of the data retrieval process. In this paper, we introduce several new trace generation models and open questions relevant to trace reconstruction for immunogenomics and DNA data storage, survey theoretical results on trace reconstruction, and highlight their connections to computational biology. Throughout, we discuss the applicability and shortcomings of known solutions and suggest future research directions.
\end{abstract}


%
\IEEEpeerreviewmaketitle

\section{Introduction}
%
%
%
%
\IEEEPARstart{T}{wo}  decades ago, Vladimir Levenshtein introduced the Trace Reconstruction Problem, reconstructing an unknown \textit{seed} string from a set of its error-prone copies, which are referred to as \textit{traces}~\cite{levenshtein1997reconstruction}. 
In information-theoretic terminology, the seed string is observed by passing it through a noisy channel multiple times. Levenshtein set forth the challenge of developing efficient algorithms to infer the seed string and characterizing the number of traces needed for its reconstruction~\cite{levenshtein2001efficient_IT, levenshtein2001efficient_JCTA}. He succeeded in solving these problems in the case of the \textit{substitution channel}, where random symbols in the seed string are mutated independently, and demonstrated that a small number of deletions or insertions may be tolerated. 
A few years later, Batu et al.~\cite{BatuKannan04-RandomCase} analyzed the trace reconstruction problem in the \textit{deletion channel}, where random symbols are deleted from the seed string independently so that a trace is a random subsequence of the seed string. 

After these seminal papers~\cite{levenshtein1997reconstruction, levenshtein2001efficient_IT, levenshtein2001efficient_JCTA, BatuKannan04-RandomCase}, trace reconstruction has received a lot of attention, especially in the last few years~\cite{abroshan2019coding, brakensiek2019coded, de2019optimal, Chase19, CGMR, HartungHP18, holden2020lower, HolensteinMPW08,  holden2018subpolynomial, KMMP19, magner2016fundamental, mao2018models, tr-revisited, nazarov2017trace, srinivasavaradhan2018maximum, viswanathan}. 
However, despite a wealth of theoretical work, there is a surprising lack of practical trace reconstruction algorithms. 
Although Batu et al.,~\cite{BatuKannan04-RandomCase} and many follow-up studies motivated trace reconstruction by the \textit{multiple alignment problem} in computational biology~\cite{compeau2015bioinformatics}, we are not aware of any software tools that use trace reconstruction for constructing multiple alignments and applying them for follow-up biological analysis. 

Transforming a biological problem into a well-defined algorithmic problem comes with many challenges. An attempt to model all aspects of a biological problem often results in an intractable algorithmic problem while ignoring some of its aspects (like in the initial formulation of the Trace Reconstruction Problem) may lead to a solution that is inadequate for practical applications. Computational biologists try to find a balance between these two extremes and typically use a simplified (albeit inadequate) problem formulation to develop algorithmic ideas that eventually lead to practical (albeit approximate) solutions of a more complex biological problem.

The first applications of trace reconstruction emerged only recently in the context of two rapidly developing research areas: personalized immunogenomics \cite{safonova2019novo, bhardwaj2019mining_d } and DNA data storage~\cite{ceze2019molecular,church2012next, bornholt2016dna, erlich2017dna, goldman,meiser2020reading,organick2018random,shipman2017crispr,yazdi2015dna, yazdi2017portable}. In this survey paper, we identify a variety of open trace reconstruction problems motivated by immunogenomics and DNA data storage, describe several practically motivated objectives for trace reconstruction, and discuss the applicability and shortcomings of known solutions. Our goal is to introduce information theory experts to emerging practical applications of trace reconstruction, and, at the same time, introduce computational biology experts to recent theoretical results in trace reconstruction. 

\subsection{Trace Reconstruction in Computational Immunology}
\subsubsection*{\textbf{How have we survived an evolutionary arms race with pathogens?}}
Humans are constantly attacked by pathogens that reproduce at a much faster rate than humans do. How have we survived an evolutionary arms race with pathogens that evolve a thousand times faster than us? 

All vertebrates have an \textit{adaptive immune system} that uses the \textit{VDJ recombination} to develop a defensive response against pathogens at the time-scale at which they evolve. It generates a virtually unlimited variety of \textit{antibodies}, proteins that recognize a specific foreign agent (called \textit{antigen}), bind to it, and eventually neutralize it. There are $\approx10^8$ antibodies circulating in a human body at any given moment (unique for each individual!) and this set of antibodies is constantly changing. How can a human genome (only $\approx20,000$ genes) generate such a diverse defense system? 

\subsubsection*{\textbf{VDJ recombination}}
\begin{figure*}[ht]
\centering
\includegraphics[width = 0.8\textwidth]{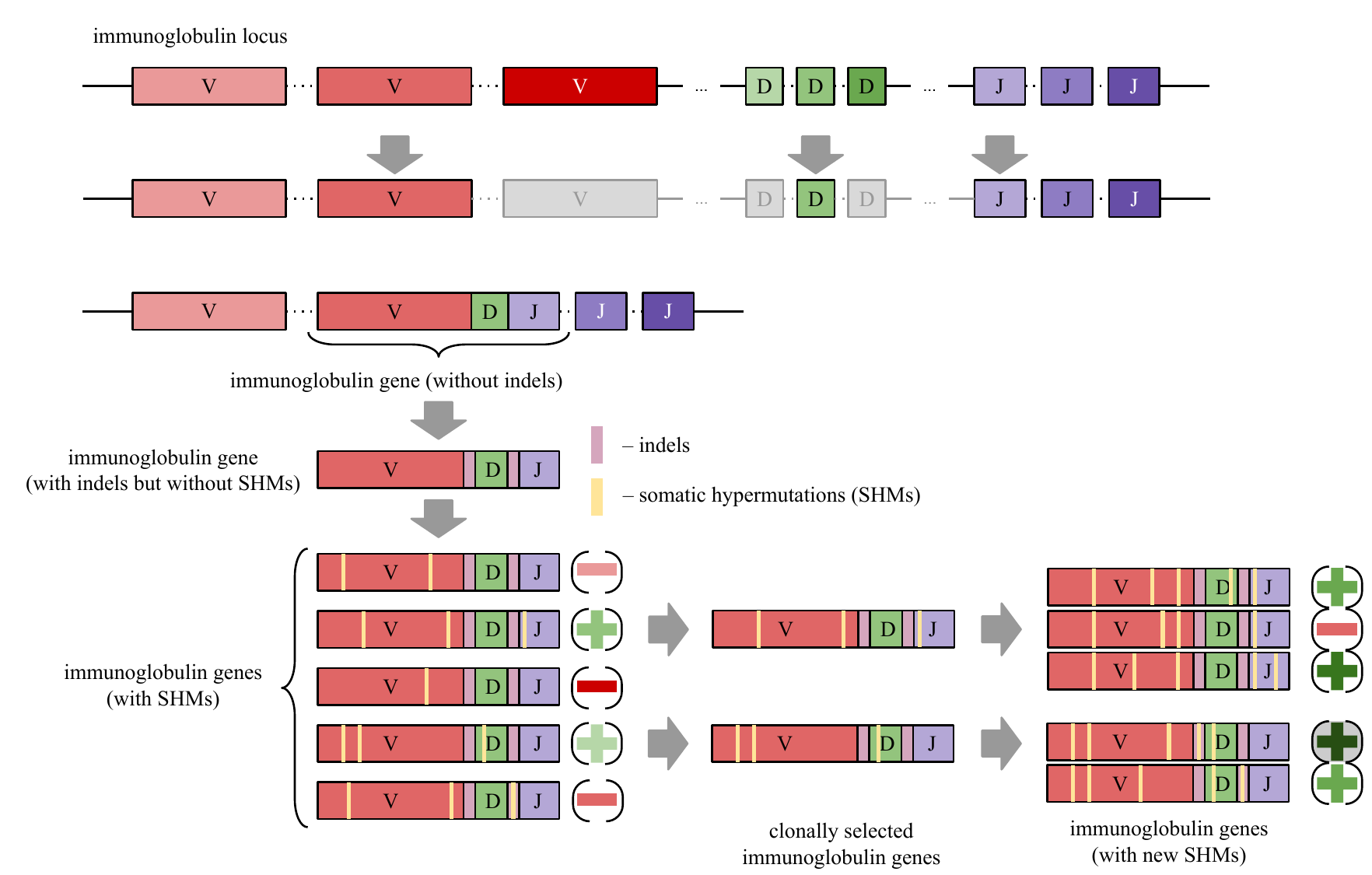}
\caption{ Generation of an antibody repertoire. The VDJ recombination affects the immunoglobulin locus that includes three sets of genes: V (variable), D (diversity), and J (joining). It randomly selects one gene from each set and concatenates them. The resulting sequence represents a potential immunoglobulin gene that encodes an antibody. However, this simple representation of an immunoglobulin gene is unrealistic since real immunoglobulin genes have indels at the V-D and D-J junctions. Somatic hypermutations (SHMs) further change the sequence of an immunoglobulin gene and thus affect its affinity. While some mutations increase affinity (sequences marked by the green `+' signs), other mutations reduce it (sequences marked by the red `-' signs). The clonal selection process iteratively retains antibodies with increased affinities and filters out antibodies with reduced affinities, thus launching an evolutionary process that eventually generates a high-affinity antibody able to neutralize an antigen (an antibody marked by a circled dark green `+' sign).   }
\label{fig_vdj_recomb}
\end{figure*}

In 1987, Susumu Tonegawa received the Nobel Prize for the discovery of the VDJ recombination~\cite{delves2017essential}. 
The \textit{immunoglobulin locus} is a 1.25 million-nucleotide long region in the human genome that contains three sets of short segments known as \textit{V, D,} and \textit{J genes} (40 V, 27 D, and 6 J genes). Figure \ref{fig_vdj_recomb} illustrates the VDJ recombination process that selects one V gene, one D gene, and one J gene and concatenates them, thus generating an \textit{immunoglobulin gene} that encodes an antibody. 
In our discussion, we hide some details to make the paper accessible to information theorists without immunology background. For example, although there are multiple immunoglobulin loci in the human genome, we limit attention to the 1.25 million-nucleotide long \textit{immunoglobulin heavy chain locus}. Although we stated above that an immunoglobulin gene encodes an antibody, in reality it encodes only the \textit{heavy chain} of an antibody (antibodies are formed by both heavy and light chains).   

Since the described process can generate only $40\times27\times6 = 6480$ antibodies, it cannot explain the astonishing diversity of human antibodies. However, the VDJ recombination is more complex than this: it deletes some nucleotides at the start and/or the end of V, D, J genes and inserts short stretches of randomly generated nucleotides (\textit{non-genomic insertions}) between V-D and D-J junctions. Such \textit{insertions and deletions} (\textit{indels}) greatly increase the diversity of antibodies generated through the VDJ recombination process. But this is only the beginning of the molecular process that further diversifies the set of antibodies. 

\subsubsection*{\textbf{Somatic hypermutations and clonal selection}}
Indels greatly increase the diversity of antibodies but even this diversity is insufficient for neutralizing a myriad of antigens that the organism might face. However, the VDJ recombination generates sufficient diversity to achieve an important goal---some of the generated antibodies in this huge collection bind to a specific antigen, albeit with low \textit{affinity} (i.e., the strength of antibody-antigen binding) that is insufficient for neutralizing the antigen. The adaptive immune system uses an ingenious evolutionary mechanism for gradually increasing the affinity of binding antibodies and thus eventually neutralizing an antigen~\cite{delves2017essential}.

Once an antibody binds to an antigen (even an antibody with a low affinity), the corresponding immunoglobulin gene undergoes random mutations (referred to as \textit{somatic hypermutations} or \textit{SHMs}) that can both increase and reduce the affinity of an antibody. To enrich the pool of antibodies with high affinity, these mutations are iteratively accompanied by the \textit{clonal selection} process that eliminates antibodies with low affinity (Figure \ref{fig_vdj_recomb}). The iterative somatic hypermutations and clonal selection are not unlike an extremely fast evolutionary process that generates a huge variety of antibodies from a single initial antibody and eventually leads to generating a new high-affinity antibody able to neutralize an antigen. 

 \subsubsection*{\textbf{Personalized immunogenomics}}
Modern DNA sequencing technologies sample the set of antibodies by generating sequences of millions of randomly selected immunoglobulin genes (\textit{antibody repertoire}) out of  $\approx10^8$ distinct antibodies circulating in our body. Analysis of antibody repertoires across various patients opens new horizons for developing  antibody-based drugs, designing vaccines, and finding associations between genomic variations in the immunoglobulin loci and diseases. The emergence of antibody repertoire datasets in the last decade raised new algorithmic problems that remain largely unsolved. 

The immunoglobulin locus is a highly variable region of the human genome---the sets of V, D, and J genes (referred to as \textit{germline genes}) differ from individual to  individual. 
Identifying germline V, D, and J genes in an individual is important since variations in these genes have been linked to various diseases \cite{watson2012immunoglobulin}, differential response to infection, vaccination, and drugs~\cite{parameswaran2013convergent}, and disease susceptibility~\cite{watson2012immunoglobulin, avnir2016ighv1}. The ImMunoGeneTics (IMGT) database of variations in germline genes remains incomplete even in the case of well-studied human genes~\cite{collins2015mouse}. In the case of immunologically important model organisms, such as camels or sharks, the germline genes remain largely unknown. Unfortunately, since assembling the sequence of the highly repetitive immunoglobulin locus faces challenges \cite{luo2019worldwide} and does not provide one with information on how various germline genes contribute to an antibody repertoire, the efforts like the 1,000 Genomes Project have resulted only in limited progress toward inferring the population-wide census of human germline genes~\cite{yu2017database, watson2017comment}.


Since the information about germline genes in an individual (personalized immunogenomics data) is typically unavailable, researchers use the reference genes instead of personal germline genes, thus limiting various immunogenomics applications. Personalized immunogenomics studies attempt to derive the germline genes by analyzing antibody repertoires. Each antibody can be viewed as a trace generated from the three sets of unknown seed strings (all V genes, all D genes, and all J genes in an individual) through the VDJ recombination and somatic hypermutations (Figure \ref{fig_vdj_recomb}). Hence, one can reformulate reconstruction of germline genes from an antibody repertoire as a novel Trace Reconstruction Problem. In Section \ref{sec_immunology_1}, we describe a series of problems with gradually increasing complexity that model antibody generation from the germline genes.

\subsection{Trace Reconstuction in DNA Data Storage}
\begin{figure*}[t]
\centering
\includegraphics[width = 0.85\textwidth]{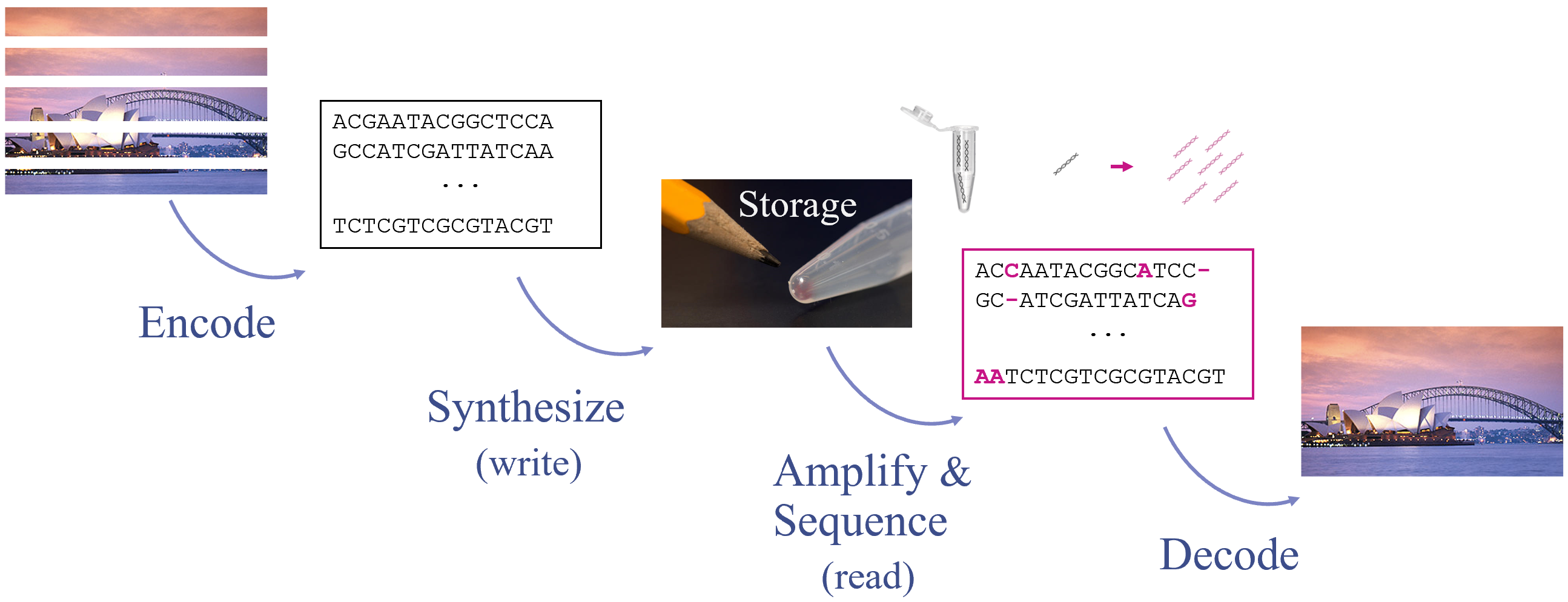}
\caption{The DNA data storage and retrieval pipeline. Trace reconstruction problems come into play just before the Decode step.}
\label{fig_dna_storage}
\end{figure*}

DNA has emerged as a potentially viable storage medium for large quantities of digital data~\cite{ceze2019molecular,church2012next, bornholt2016dna, erlich2017dna, goldman,meiser2020reading,organick2018random,shipman2017crispr,yazdi2015dna, yazdi2017portable}. A digital file can be encoded by a collection of DNA sequences where each individual sequence represents a small part of the data. One application is archival storage, where DNA promises to have orders of magnitude improved data density and durability as compared to existing storage media (e.g., magnetic tapes or solid state). The field is rapidly growing, and current DNA data storage systems can store and retrieve hundreds of megabytes of data, with many additional features such as random data access~\cite{organick2018random}. We provide an overview of DNA data storage and highlight the role that trace reconstruction plays in the data retrieval process~\cite{yazdi2017portable, organick2018random}. Figure \ref{fig_dna_storage} depicts the core components of the storage and retrieval pipeline.

\subsubsection*{\textbf{Storing the data}}
Storing a file in DNA involves several steps. First, the digital file is compressed and partitioned into small, non-overlapping blocks. Then, each individual block is either encoded using an error-correcting code or is randomized using an independent pseudo-random sequence. This provides a set of strings that encode the content of the digital file. To store the location of each block, an address is added to each string. Finally, a global error-correcting code is applied to the resulting set of strings, and the strings are translated into the $\{A, C, G, T\}$ alphabet. If multiple files are stored together, then a file identifier is also added in the form of a \textit{DNA primer} (a short nucleotide string). This process results in a large collection of short strings (for example, millions of strings, each containing hundreds of characters). This set of strings, which we call {\em seed strings}, are then synthesized into real DNA molecules and stored in a tube until the file is ready to be retrieved. The synthesis process generates many copies of each seed string. 

\subsubsection*{\textbf{Retrieving the data}}
The stored information is read using standard DNA sequencing machines such as DNA sequencers produced by Illumina. A small amount of DNA is extracted from a tube so that the remaining DNA may be used for other retrieval attempts later on. Since this amount may be insufficient for reading DNA (sequencing machines have limitations with respect to the minimal amount of DNA they can sequence), the extracted DNA is amplified using \textit{polymerase chain reaction} (\textit{PCR}) to generate multiple copies (e.g., 5--30) of each DNA molecule in a sample. The PCR step enables random access retrieval---to access a subset of files, it suffices to copy and sequence the subset of seed strings with \textit{primers} (file IDs) corresponding to these files. 

However, the PCR process introduces additional errors in each of the amplified copies. Since DNA sequencing machines are not able to identify the error-free sequence of nucleotides in a DNA molecule, they add extra errors to the previously introduced amplification errors. The combination of amplification and sequencing errors typically results in $\approx 1$--$2\%$ error rate (substitutions and indels). However, there is some debate about the rate and the most common type of error~\cite{potapov2017examining, sabary2019solqc}. The output of sequencing is a set of strings that contains several error-prone copies (called {\em reads}) of each originally synthesized seed string.
Much longer seed strings (tens of thousands of nucleotides versus hundreds of nucleotides in existing applications) can be sequenced using the recently emerged \textit{long-read sequencing technologies}  but the current error rate of such technologies is $\approx 10 \%$, with a large proportion of indels~\cite{ yazdi2017portable, organick2018random, chandak2019, lopez2019dna}. 

\subsubsection*{\textbf{DNA data retrieval as a trace reconstruction problem}}
After the sequencing reads are generated, the goal is to recover the seed strings from the observed error-prone reads that have indels and substitutions. The first challenge is to determine which reads correspond to which seed strings by clustering reads so that each cluster contains the error-prone copies derived from a single seed string~\cite{rashtchian2017clustering}. In some DNA data storage systems, the seed strings are randomized or encoded in a such a way that they have large pairwise edit distance~\cite{GMR16, navarro2001guided, organick2018random, schimd2019bounds}. This property simplifies the clustering problem because the underlying clusters are well-separated. In this context, clustering algorithms have been developed that scale to billions of reads~\cite{rashtchian2017clustering}.

Recovering the seed strings from the reads can be formulated as a trace reconstruction problem. Each seed string is observed a small number of times, where the error-prone copies (traces) correspond to the reads in a cluster. The objective is to recover as many seed strings as possible. A small number of missing or erroneous seed strings may be tolerated because of the error correction methods. Consequently, it suffices to ensure that a reconstruction algorithm recovers a seed string with probability \textit{ReconstructionRate}, where the exact success probability depends on the amount of redundancy in the error-correcting code (e.g., the default value may be \textit{ReconstructionRate} $ =0.95$). There is a trade-off where having more traces leads to lower error rate in reconstruction, but it incurs a higher sequencing cost and time. In practice, it is typical to use clusters that contain 5--30 reads (traces)~\cite{organick2018random}.

While we focus on trace reconstruction problems in DNA data storage, there are many other challenges and recent results, including better automation methods~\cite{newman2019high, willsey2019puddle}, alternative synthesis schemes~\cite{anavy2019data, tabatabaei2020dna}, improved density and robustness using codes~\cite{chandak2019,dube2019dna,erlich2017dna, fei2019ldpc,gabrys2020mass,jain2020coding, immink2017design, lenz2019anchor, lenz2020coding, lenz2018coding,lenz2020covering, lenz2019upper, pattabiraman2020coding, sima2019coding, shinkar2019clustering}, and more realistic error models and fundamental limits~\cite{conde2018nanopore, heckel2019characterization,heckel2017fundamental, mao2018models, organick2020probing}. For more details about DNA data storage, see the following surveys and references therein~\cite{ceze2019molecular, yazdi2015dna}.

\subsection{Similarities and differences of the two applications}
The trace reconstruction problems for immunogenomics and DNA data storage are distinct, both in terms of the trace generation models and how well the models have been studied in the literature.

In immunogenomics, the traces contain important mutations that are introduced during the VDJ recombination and somatic hypermutagenesis. While sequencing and amplification technologies also introduce errors in sequenced antibody repertoires, their rate is much lower compared to the mutations introduced at the antibody generation step. Therefore, we ignore sequencing and amplification errors in immunogenomics applications and focus on the mutations. In contrast, the errors in the DNA data storage applications are only because of the artifacts of the process used to access the data stored in DNA and thus cannot be ignored. 

The seed strings in immunogenomics represent real genetic data, whereas the DNA data storage sequences are synthesized to represent information in a digital file. While there has been a considerable amount of work in trace reconstruction problems motivated by DNA data storage, trace reconstruction studies in immunogenomics have only started to emerge~\cite{safonova2019novo, bhardwaj2019mining_d}. 

\subsection{Outline}
The rest of the paper is organized as follows. \newtext{In Section \ref{sec_algo_formulations}, we introduce the algorithmic and information-theoretic formulations of trace reconstruction.}
Section \ref{sec_immunology_1} describes trace generation in computational immunogenomics. In Sections \ref{sec_immunology_simple} and \ref{sec_immunology_adequate}, we introduce the D genes trace reconstruction problem. In Section \ref{sec_immunology_vdj_reconstruction}, we introduce a more complex problem of reconstructing V, D, and J genes that are concatenated together to form antibodies. Section~\ref{sec_dna_storage} describes the theoretical formulation of  trace reconstruction problems for DNA data storage. In  Section~\ref{sec_theoretical_results}, we survey  theoretical results  and practical solutions to the trace reconstruction problem for the deletion channel, along with open problems relevant to developing DNA data storage. \newtext{Finally, in Section~\ref{sec_conclusion} we propose several directions for future work.}

\section{Algorithmic and information-theoretic formulations}
\label{sec_algo_formulations}

In this section, we formalize the algorithmic goals of the trace reconstruction problems. We begin by considering an abstract model, where a single, unknown seed string $s$ generates a random trace $c$ with probability $\mathbf{Pr}(c \mid s)$. For each possible trace $c$ and seed string $s$, the model specifies  $\mathbf{Pr}(c \mid s)$. To recover the seed string $s$, the reconstruction algorithm receives a collection of traces generated from $s$, which we refer to as a {\em trace-set} $C = \{c_1, c_2, \ldots, c_T\}$. For simplicity, we assume that the traces are independent and identically distributed, and hence, $$\mathbf{Pr}(C \mid s) = \prod_{i=1}^T  \mathbf{Pr}(c_i \mid s).$$  Given an integer $T$, we use $\mathcal{C}_{T}$ to denote the collection of all possible sets of~$T$ strings over a fixed alphabet, and we note that $\sum_{C \in \mathcal{C}_{T}} \mathbf{Pr}(C \mid s)=1$.

We also consider cases where the generation process involves sets of seed strings. In these cases, one string is sampled at a time from a set, and the traces are independently generated from the sampled strings (sometimes concatenating groups of traces to obtain the final trace-set). For example, we can consider the two step process where a seed string $s$ is uniformly randomly selected from an unknown {\em seed-set} $S$, and then $s$ generates a trace. Given a seed-set $S = \{s_1,\ldots, s_M\}$, the probability of a trace-set $C$ is
$$\mathbf{Pr}(C \mid S)  = \prod_{i=1}^T \mathbf{Pr}(c_i \mid S) = \prod_{i=1}^T \left(\frac{1}{M} \sum_{j = 1}^M \mathbf{Pr}(c_i \mid s_j) \right).$$
In other words, in the multiple seed string case, we can still define the probability of a trace-set $C$ in terms of the probability of generating a single trace from a single seed string. The goal is to recover all or most of the strings in $S$ by using a trace-set generated via this two step process.

\newtext{
We next discuss how to evaluate a reconstruction algorithm $A$ that takes as input the trace-set $C$ and outputs a string $A(C)$. We assume that the algorithm knows the trace generation model, that is, for any trace $c$ and seed string $s$, it knows the probability $\mathbf{Pr}(c \mid s)$. The goal is to reconstruct the seed string using the traces. 
The fact that the traces themselves are random means that there are at least two ways to evaluate a reconstruction algorithm. 

The {\em maximum likelihood estimate (MLE)} is a string $\widehat s$ that maximizes $\mathbf{Pr}(C \mid \widehat s)$ among all seeds.
As the probabilities are known to the algorithm, the MLE can always be computed by exploring all strings (i.e., brute-force search) as long as the set of possible candidate strings is finite. The trace generation models that we consider have the property that the maximum length of the seed string can be inferred from the trace-set with high probability. Therefore, the brute-force search can be taken over a finite set of strings. For some models, an efficient algorithm computing the MLE is known, with running time that is polynomial in the number $T$ of traces and the length $|s|$ of the seed string (see Section~\ref{sec_immunology_simple}). However, for many trace generation models, computing the MLE in polynomial time is currently an open question (i.e., the only known solution is brute-force search). 

To circumvent the difficulty of the maximum likelihood objective, previous work instead measures the probability that an algorithm outputs the seed string $s$ used to generate the traces. The trace-set is viewed as a random input, and the probability is taken over the randomness in the trace generation process. We start with definitions for a fixed, but unknown, seed string $s$, and we later also consider $s$ itself being random. Define the {\em success probability} of an algorithm~$A$ and a seed string $s$ as 
$$
P_A(s, T) = \sum_{C \in \mathcal{C}_{T}} \mathbf{Pr}(C \mid s) \cdot 1_{ \{A(C) = s\}  },
$$
where $1_{\{A(C) = s\}}$ is the indicator function for the event $\{A(C) = s\}$ that the algorithm outputs the seed string $s$. It is straightforward to extend the definition of $P_A(s, T)$ to randomized algorithms; the output $A(C)$ would also be a random variable, and the term $1_{ \{A(C) = s\} }$  would be replaced with $\mathbf{Pr}(A(C) = s)$.

Let $\U$ be a universe of possible seed strings (e.g., all strings of a certain length over a binary or quaternary alphabet). We define the {\em worst-case success probability} of algorithm~$A$ for trace-sets of size~$T$ over universe~$\U$ as
$$
P_A(\U, T)  = \min_{s \in \U} P_A(s, T).
$$
Then, the {\em worst-case} trace reconstruction problem is to develop an efficient algorithm that maximizes $P_A(\U, T)$. The definition above guarantees that the algorithm succeeds with probability at least $P_A(\U, T)$ when $s$ is an arbitrary seed string from the universe and the trace-set has size $T$.

We also consider the {\em average-case} trace reconstruction problem,
where the seed string $s$ is chosen uniformly at random from the universe (instead of being arbitrary, as in the worst-case version). More precisely, the goal is to develop an efficient algorithm $A$ that maximizes the  
{\em average-case success probability}, which is defined as 
$$
\widetilde P_A(\U, T)  = \frac{1}{|\U|} \sum_{s\in\U} P_A(s, T).
$$
Notice that the probability here is taken over both the seed string $s$ and the trace-set $C$. 
The average-case formulation leads to a nice connection to the MLE. Expanding $P_A(s, T)$, we have that

\begin{IEEEeqnarray*}{rCl}
\widetilde P_A(\U, T) & = & \frac{1}{|\U|} \sum_{s\in\U} \sum_{C \in \mathcal{C}_{T}} \mathbf{Pr}(C \mid s) \cdot 1_{ \{A(C) = s\}  }\\
& = &  \sum_{C \in \mathcal{C}_{T}} \frac{1}{|\U|} \sum_{s\in\U}  \mathbf{Pr}(C \mid s) \cdot 1_{ \{A(C) = s\}  } 
\end{IEEEeqnarray*}

Therefore, the inner sum over $s \in \U$ is maximized when $A$ outputs the string $\widehat s$ maximizing $\mathbf{Pr}(C \mid \widehat s)$, or in other words, when the algorithm outputs the MLE. 

We note that algorithm does not know the seed string, and hence, it cannot determine whether it outputs $s$ or some other string $s'$ that could have generated the trace-set. In contrast, the MLE is always rigorously defined because it allows the algorithm to output any string that maximizes the likelihood. To rigorously reason about the maximum success probability formulation, we assume that the trace-set is large enough so that a unique seed string must have generated the traces with high probability, and hence, the algorithm can recover this string with high probability. Later on, we also discuss how to empirically determine the success probability with a benchmark dataset.  

In summary, when the seed string is random (i.e., the average-case version), then the maximum likelihood solution also maximizes the success probability $\widetilde P_A(\U, T)$. In particular, the ideal solution to the average-case trace reconstruction problem would be an efficient algorithm that computes the MLE, with running time that is polynomial in  the number of traces and the seed string length. For the new models that we introduce, we remain hopeful that such an algorithm can be found. However, the only presently known algorithm for all but one of the models is to perform brute-force search.  Moreover, in the worst-case version, the MLE may not maximize the success probability $P_A(s, T)$, and these two formulations may lead to different optimal algorithms.

In Section~\ref{sec_immunology_1}, we introduce various trace generation models in computational immunogenomics. For each model, we provide a problem statement that asks for an MLE solution, i.e., an algorithm that outputs a seed string (or a seed-set) that maximizes the likelihood of a given trace-set. However, we also note that it would be valuable to develop an algorithm with high success probability when the input is viewed as a random trace-set. While both of these are valid and important formulations, the MLE version is a long-standing tradition in bioinformatics that is widely used in such areas as computing phylogenetic trees~\cite{guindon2010new} and genetic linkage analysis~\cite{bailey2011linkage}. In immunogenomics, MLE was used for computing antibody clonal trees~\cite{hoehn2017phylogenetic}, modeling VDJ recombination~\cite{murugan2012statistical,ralph2016likelihood}, and modeling antibody-antigen interactions~\cite{pan2011selective,watabe2007likelihood}. 
On the other hand, information theory and computer science researchers may prefer to develop (approximation) algorithms that are evaluated based on their success probability. Thus, we briefly discuss evaluation metrics before introducing the models.
}
\subsection{Approximation Algorithms and Empirical Success Probability}
\label{approx_overview}

As for many other bioinformatics problems, since brute-force solutions are prohibitively slow, the goal is to develop fast approximation or heuristic algorithms that are practical for typical input sizes. For an analogy, although the edit distance problem between two sequences  can be solved in polynomial time~\cite{Lev65}, the closely related \textit{sequence alignment} problem between multiple sequences is NP-hard \cite{wang1994complexity}. Nevertheless, since the multiple sequence alignment problem is at the heart of sequence comparison in bioinformatics, hundreds of heuristic algorithms have been developed for solving it \cite{notredame2007recent}. The ultimate goal of these algorithms is to generate biological insights, and hence, they are often benchmarked on datasets with known solutions \cite{thompson2011comprehensive}. 

Turning back to trace reconstruction problems, it would often suffice to output the MLE on most trace-sets, instead of all of them (e.g., failing with vanishingly small probability). Alternatively, when it is difficult to find the entire seed string $s$ maximizing $\pCs$, it may be possible to find a sufficiently long substring instead. Doing so could further enable finding the entire seed string through a complementary experimental approach. For example, a seed string reconstructed by an approximation or heuristic algorithm can be later validated and error-corrected by using genomics data that complements the immunogenomics data~\cite{bhardwaj2019mining_d}.

We also mention one more choice: is the number of traces fixed in advance or not? For a fixed number of traces $T$, the goal is to design an algorithm with highest possible success probability. Alternatively, since the success probability increases as $T$ increases, we consider an additional input parameter \textit{ReconstructionRate}, where $0 \leq $ \textit{ReconstructionRate} $\leq 1$ and the goal is to design an algorithm with success probability surpassing \textit{ReconstructionRate} using as few traces as possible. Formally, we want to determine the minimum value $T^*$ such that the trace reconstruction problem with $T$ traces is feasible for a given \textit{ReconstructionRate} as long as $T \geq T^*$. This value $T^*$ is called the trace complexity, and we discuss it further in Section~\ref{sec_dna_storage}.  We also note that the success probability can be driven to one by using more traces, assuming it starts above 0.5. Indeed, taking the majority vote over $O(\log (1/\beta))$ trials for any value $0 < \beta < 1$ will lead to success probability $1-\beta$, which follows via a Chernoff bound. Both algorithmic formulations are relevant for practical applications.

For the immunology models, we consider a fixed number of traces. The reason is that the number of traces depends on multiple factors---such as the reconstruction of \textit{clonal trees} during antibody development~\cite{yermanos2018tracing} or selecting the best candidate for follow-up antibody engineering efforts~\cite{hsiao2019immune}---and accurate reconstruction of germline genes is only one of them. For the DNA data storage models, we consider an information-theoretical perspective and focus on determining the minimum number of traces that suffice for a certain success probability. 

The average-case success probability can be empirically calculated by choosing the seed string $s$ at random and testing whether $A(C) = s$ when the trace-set is generated at random from~$s$. For the worst-case success probability, it is infeasible to compute the minimum over all possible length $n$ strings. Instead, it would be easier to use seed strings from a benchmark dataset. For example, if the \textit{ReconstructionRate} is 0.95, then the algorithm will likely output $A(C) = s$ at least 95 times over 100 randomly generated trace-sets, and this should hold for each seed string $s$ from the dataset. In the DNA data storage application, the seed strings are constructed synthetically during the storage process, and therefore, they may be used as a benchmark.

\section{Trace Generation in Computational Immunogenomics}
\label{sec_immunology_1}

\subsubsection*{\textbf{Reconstructing D genes is more difficult than reconstructing V and J genes}}
Inferring the sequences of germline genes using immunosequencing data obtained from an individual antibody repertoire is an important problem\cite{boyd2010individual, gadala2015automated,corcoran2016production,zhang2016impre,safonova2019novo, bhardwaj2019mining_d}. In the case of V and J genes, this challenge was addressed by~\cite{gadala2015automated,corcoran2016production,zhang2016impre, ralph2016consistency}.  Reconstruction of shorter D genes is a more challenging task~\cite{ralph2016consistency}.  D genes contribute to the \textit{complementarity determining region 3} (\textit{CDR3}) that spans the V-D and D-J junctions and represents an important and highly divergent part of antibodies that accumulates many SHMs. Since D genes typically get truncated on both sides during VDJ recombination, the CDR3 typically contains a truncated D gene. Each CDR3 also contains some random insertions at the V-D and D-J junctions.  
These truncations and insertions, combined with the fact that D genes are much shorter than V and J  genes, make the task of aligning various CDR3s (and thus aligning segments of D genes that survive within these CDR3s) more difficult than alignment of longer and typically less mutated fragments of immunoglobulins that originated from V and J genes.

The biologically adequate problem formulations in immunogenomics are rather complex, making it difficult to develop and test algorithmic ideas for solving these problems. That is why the usual path toward solving such problems is to start from simple and often inadequate formulations that however shed light on algorithmic ideas that can be used for solving more complex problems \cite{medvedev2019modeling}. We follow this path by starting with a simple formulation for the problem of inferring D genes from CDR3s extracted from an antibody repertoire. Although efficient algorithms for the complex biologically adequate problems remain unknown, the recently developed MINING-D heuristic \cite{bhardwaj2019mining_d} led to the discovery of previously unknown D genes across multiple species. After describing open problems relevant to finding new D genes, we formulate more difficult problems relevant to inferring the sets of V, D, and J genes (rather than D genes only).

\subsubsection*{\textbf{Generating CDR3 from a D gene}}
We denote the length of a string $s$ as $|s|$ and the concatenation of strings $s_1$ and $s_2$ as $s_1*s_2$. We refer to a random string of length $l$ (each symbol is generated uniformly at random from a fixed alphabet $\caliA$) as $r^l$. Given an integer $t$, we define a random string ${Random}_{\leq t}$ as ${r}^l$, where an integer $l$ is sampled uniformly at random from $[0, t]$. In this paper, $\caliA = \{A, G, C, T\}$.

\begin{figure}[t]
\centering
\includegraphics[width=\columnwidth]{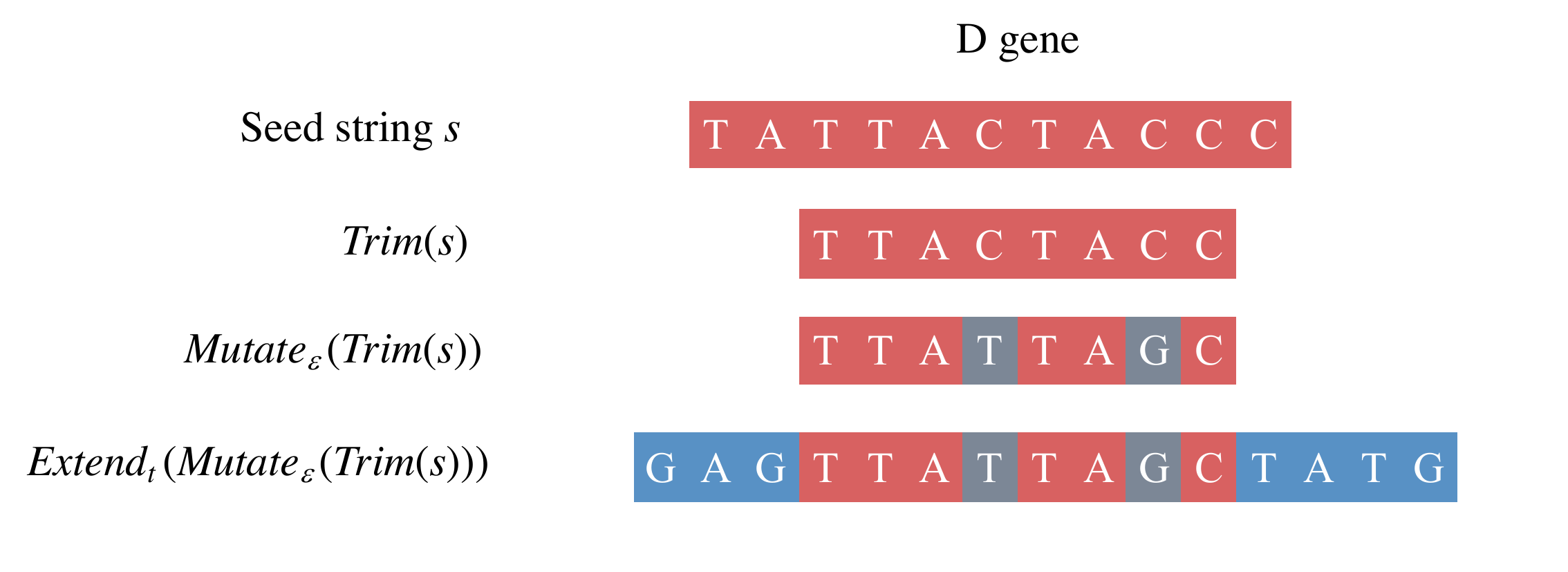}
\caption{Trim, Mutate, and Extend operations model the process of generating a CDR3 of an immunoglobulin gene from a D gene using somatic hypermutations (shown in green) and random insertions (shown in blue).}
\label{fig_trim_mutate_extend_model}
\end{figure}

Below we describe various models for generating traces from a seed string or from a seed-set. In all models, we assume that each trace is generated independently.  
To model generation of a CDR3 (trace) from a D gene (seed) in the models below, we describe the following operations on a string~$s$ (Figure \ref{fig_trim_mutate_extend_model}):
\begin{itemize}
\item \textit{Trim(s)}: A pair of integers $l$ and $k$ are sampled uniformly at random from the set of all pairs of non-negative integers $(i,j)$ satisfying the condition $i+j\leq |s|$. \newtext{The prefix of length $l$ and the suffix of length $k$ of $s$ are trimmed.}
\item \textit{$\text{Mutate}_\epsilon$(s)}: Each letter in $s$ is independently mutated with probability $\epsilon$ such that mutations into all  $|\caliA| -1$ symbols (differing from the symbol in $s$) are equally likely.  
\item \textit{$\text{Extend}_t$(s)}: \newtext{a string $R_1*s*R_2$ where $R_1$ and $R_2$ are independent instances of ${Random}_{\leq t}$.  }
\end{itemize}

\begin{figure*}[t]
     \centering
     \begin{subfigure}[t]{0.3\textwidth}
         \includegraphics[width=\textwidth]{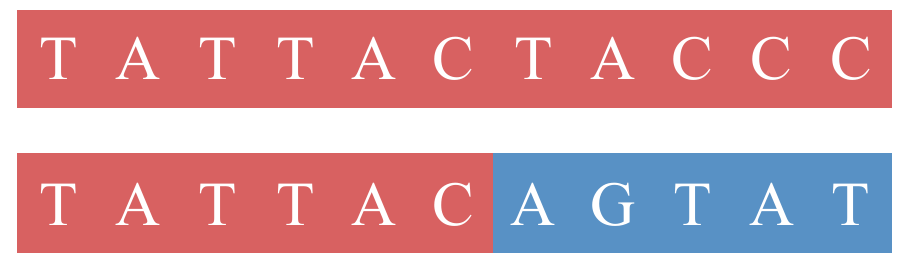}
         \caption{\textit{TrimSuffixAndExtend}}
         \label{fig_TrimSuffixAndExtend}
     \end{subfigure}\quad
     \begin{subfigure}[t]{0.3\textwidth}
         \includegraphics[width=\textwidth]{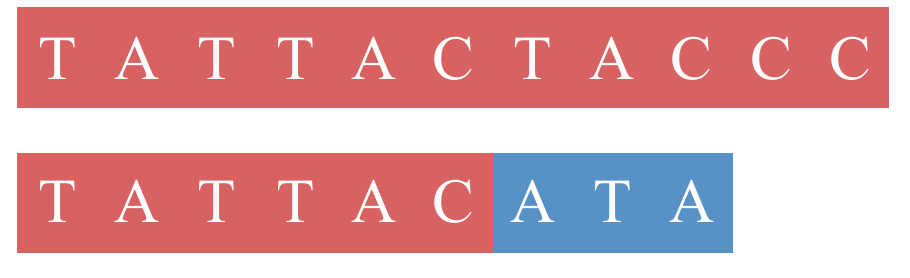}
         \caption{\textit{$\text{SuffixExtend}_t$(TrimSuffix)}}
         \label{fig_SuffixExtendTrimSuffix}
     \end{subfigure}\quad
     \begin{subfigure}[t]{0.3\textwidth}
         \includegraphics[width=\textwidth]{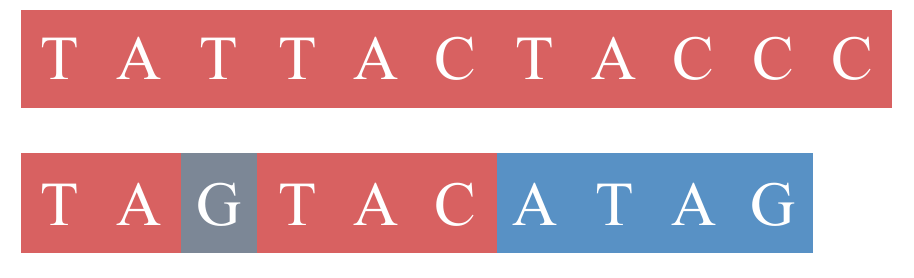}
         \caption{\textit{$\text{SuffixExtend}_t$($\text{Mutate}_{\epsilon}$(TrimSuffix))}}
         \label{fig_SuffixExtendMutateTrimSuffix}
     \end{subfigure}
     \medskip
     \vspace{.25in}
     \begin{subfigure}[t]{0.3\textwidth}
         \includegraphics[width=\textwidth]{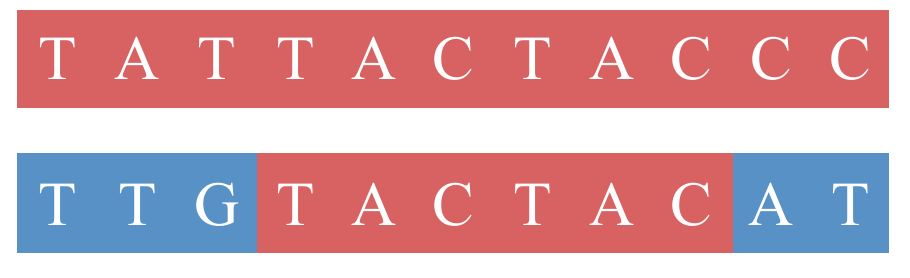}
         \caption{\textit{TrimAndExtend}}
         \label{fig_TrimAndExtend}
     \end{subfigure}\quad
     \begin{subfigure}[t]{0.3\textwidth}
         \includegraphics[width=\textwidth]{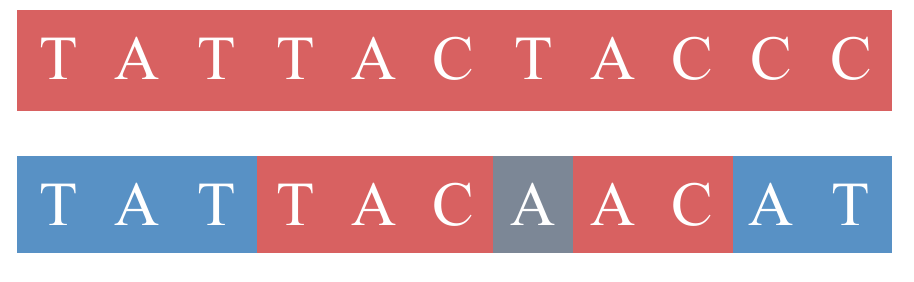}
         \caption{\textit{$\text{Mutate}_{\epsilon}$(TrimAndExtend))}}
         \label{fig_MutateTrimAndExtend}
     \end{subfigure}

        \caption{Trace generation for various trace reconstruction problems motivated by analysis of immunosequencing data. Insertions (i.e., random strings of random length) are shown in blue. Hypermutations are shown in green.}
        \label{fig_d_models}
\end{figure*}

Figure \ref{fig_trim_mutate_extend_model} illustrates the \textit{$\text{Extend}_t(\text{Mutate}_\epsilon$(Trim))} model for generating a CDR3 from a D gene using random deletions/insertions and somatic hypermutations. Before considering this rather complex model, we will consider a series of simpler (albeit less adequate) models for generating CDR3s (Figure \ref{fig_d_models}) that use the operations listed below. 

\begin{itemize}
\item \textit{TrimSuffix(s)}: an integer $k$ is sampled uniformly at random from $[0, |s|]$ and the suffix of $s$ of length $k$ is trimmed.
\item 	\textit{TrimSuffixAndExtend(s)}: an integer $k$ is sampled uniformly at random from $[0, |s|]$, the suffix of $s$ of length $k$ is trimmed, and then the resulting string is concatenated with ${r}^k$.
\item	\textit{SuffixExtend$_t$(s)}: a string $s~*$~\textit{Random}$_{\leq t}$.
\item \textit{TrimAndExtend(s)}: a pair of integers $l$ and $k$ are sampled uniformly at random from the set of all pairs of non-negative integers $(i,j)$ satisfying the condition $i+j\leq  |s|$. 
\newtext{The prefix of length $l$ and the suffix of length $k$ of $s$ are trimmed} resulting in a string \textit{Trim(s)}. \textit{TrimAndExtend(s)} is defined as ${r}^l~*$~\textit{Trim(s)}~$*~{r}^k$. 
\end{itemize}


We will start with a simple \textit{TrimSuffixAndExtend} model where the seed string and the modified strings are of equal lengths. The next  \textit{$\text{SuffixExtend}_t$(TrimSuffix)} model relaxes the assumption that the lengths of all modified strings generated from a seed string are the same since the same D gene can produce CDR3s of different lengths in the VDJ recombination process. In the next  \textit{$\text{SuffixExtend}_t$($\text{Mutate}_{\epsilon}$(TrimSuffix))} model, we further allow mutations to occur in the seed string. This is important because the immune system introduces random somatic hypermutations to increase the affinity towards an antigen. 

In the above models, only the suffix of the seed string gets trimmed in the first step. In the real VDJ recombination process, however, D genes get trimmed on both sides. To incorporate this fact in the above models, we next present the \textit{TrimAndExtend} model that allows trimming on both sides while keeping the lengths of the modified strings the same. This is analogous to the \textit{TrimSuffixAndExtend} model and the only difference between the two models is that the former gets trimmed on both sides whereas in the latter, only the suffix is trimmed. 
To introduce mutations in this model, where the seed string gets trimmed on both sides, we then present the \textit{$\text{Mutate}_{\epsilon}$(TrimAndExtend))} model, while still keeping the lengths of all modified strings the same. Finally, to allow for the possibility of different lengths of modified strings, while keeping intact the trimming from both sides and the random mutations, we introduced the \textit{$\text{Extend}_t(\text{Mutate}_{\epsilon}$(Trim))} model which is the most biologically adequate model for VDJ recombination among all introduced models.

All models presented in the next subsection can be extended to the multiple seed strings case where a seed string is chosen randomly from a seed-set, a trace is then generated from the chosen seed string according to a model, and the process is independently repeated a number of times to generate a set of traces. In Section~\ref{sec_dna_storage_multiple_seeds}, we will discuss the \textit{population recovery problem}, which also concerns reconstructing multiple seed strings under a different trace generation model.

The average length and the number of D genes varies among species---for humans and many immunologically important mammals (e.g., mice and rats), the length of D genes does not exceed 40 nucleotides and the number of D genes varies from 20 to 40. In contrast, other immunologically important mammals (e.g., cows) have long (150 nucleotides) and very repetitive D genes. Since future personalized immunogenomics studies may involve thousands or even millions of individuals, the D gene reconstruction algorithms must scale accordingly, e.g., the running time should not exceed a few hours.

\subsection{ A Simple but biologically inadequate model for D gene reconstruction}
\label{sec_immunology_simple}

\subsubsection*{\textbf{TrimSuffixAndExtend}}
Although this model (Figure \ref{fig_d_models}a) does not adequately reflect the realities of VDJ recombination, the trace reconstruction problem for this model can be efficiently solved. 
A seed string may generate the same trace for different values of the trimming integer $k$ in the \emph{TrimSuffixAndExtend} model. 
The probability $\mathbf{Pr}(c \mid s) $ that a seed string $s$ generates a trace $c$ depends only on the length~$m$ of their longest shared prefix and is given by
\begin{align*}
\mathbf{Pr}(c \mid s) &= \frac{1}{|s|+1} \sum_{k=0}^m 
\frac{1}{ {|\caliA|}^{|s| -k}} \\
&= \frac{1}{(|s|+1)({|\caliA|}^{|s|})} \times 
\frac{{|\caliA|^{m+1} -1 }}{  |\caliA| -1 }\\
& = K(|s|, |\caliA|) \times (|\caliA|^{m+1} - 1)
\end{align*}
where $K(|s|, |\caliA|)$ is constant given the length of the seed string and the alphabet size. 
The probability that a seed string~$s$ generates a trace-set $C = \{c_1, c_2, \ldots, c_T\}$ is computed as
\begin{equation}
\mathbf{Pr}(C \mid s) = \prod_{i=1}^{T} \mathbf{Pr}(c_i \mid s).
\label{eq_prod_prob}
\end{equation}

\begin{quote}
\vspace{.1in}
\textbf{Trace Reconstruction Problem in the \emph{TrimSuffixAndExtend} model}\\
\textbf{Input:} A trace-set $C$ generated from an unknown seed string according to the \emph{TrimSuffixAndExtend} model.\\
\textbf{Output:} A string $s$ maximizing $\mathbf{Pr}(C\mid s)$.
\vspace{.1in}
\end{quote}

\subsubsection*{\textbf{Solving Trace Reconstruction Problem in the TrimSuffixAndExtend model}}
$\pCs$ is maximized by one of the traces. This observation leads to an algorithm for solving the String Reconstruction Problem (with complexity $O(|s| \cdot T^2))$ that simply computes $\pCs$ for each of the $T$ traces. We describe an improved algorithm for solving this problem with a running time of $O(|s| \cdot T)$, which is linear in the input size.

Maximizing $\pCs$ is equivalent to maximizing $\prod_{i=1}^T K(|s|, |\caliA|) \times ( |{\caliA}|^{m_i +1} -1)$,  where $m_i$ is the length of the longest shared prefix between $s$ and $c_i$ \cite{bhardwaj2019mining_d}. 
Since $ K(|s|, |\caliA|) $ is a constant, it is equivalent to finding a string $s$ that maximizes
\begin{equation*}
\text{score} (C \mid s) =  \sum_{i=1}^{T} \log (|{\caliA}|^{m_i +1} -1). 
\end{equation*}

We denote $f(j) = \log (|\caliA|^{j+1} -1)$ and search for a string $s$ that maximizes $\sum_{i=1}^T f(m_i)$ where $m_i$ is the length of the longest shared prefix between $s$ and $c_i$.
We denote a $t$-symbol prefix (\emph{$t$-prefix}) of a string $c$ as $c^t$ and the set of all $t$-prefixes of strings from $C$ as $C^t$. Given a string $s$ and an integer~$t$, we say that a string $c$ is \textit{t-similar} to $s$ if $t$-prefixes of $s$ and $c$ coincide. The number of strings in $C$ that are $t$-similar to $s$ is denoted as $sim_t(C,s)$. 
Given a string $s$,

\begin{IEEEeqnarray}{rCl}
\text{score}(C^t \mid s^t) & = & \text{score}(C^{t-1} \mid s^{t-1} )  \nonumber\\
&& +\> sim_t(C, s) \times \log\left(\frac{|\caliA|^{t+1}-1}{|\caliA|^t -1}\right).
\label{eq_recurion}
\end{IEEEeqnarray}

We use this recurrence to efficiently compute  $\text{score}(C \mid s)$ for each string $s$ from $C$ using dynamic programming on a trie constructed from all traces in $C$~\cite{gusfield1997algorithms}. 
Each vertex in the trie is a $t$-prefix $s^t$ of a string from $C$, and we recursively compute $\text{score} (C^t \mid s^t)$ in each vertex of the trie using the above recurrence assuming that the score of the root is $T\cdot \log(|\caliA|-1)$.
The optimal string corresponds to the leaf node with the maximum score. 

\begin{figure*}[t]
\centering
\includegraphics[width = 0.7 \textwidth]{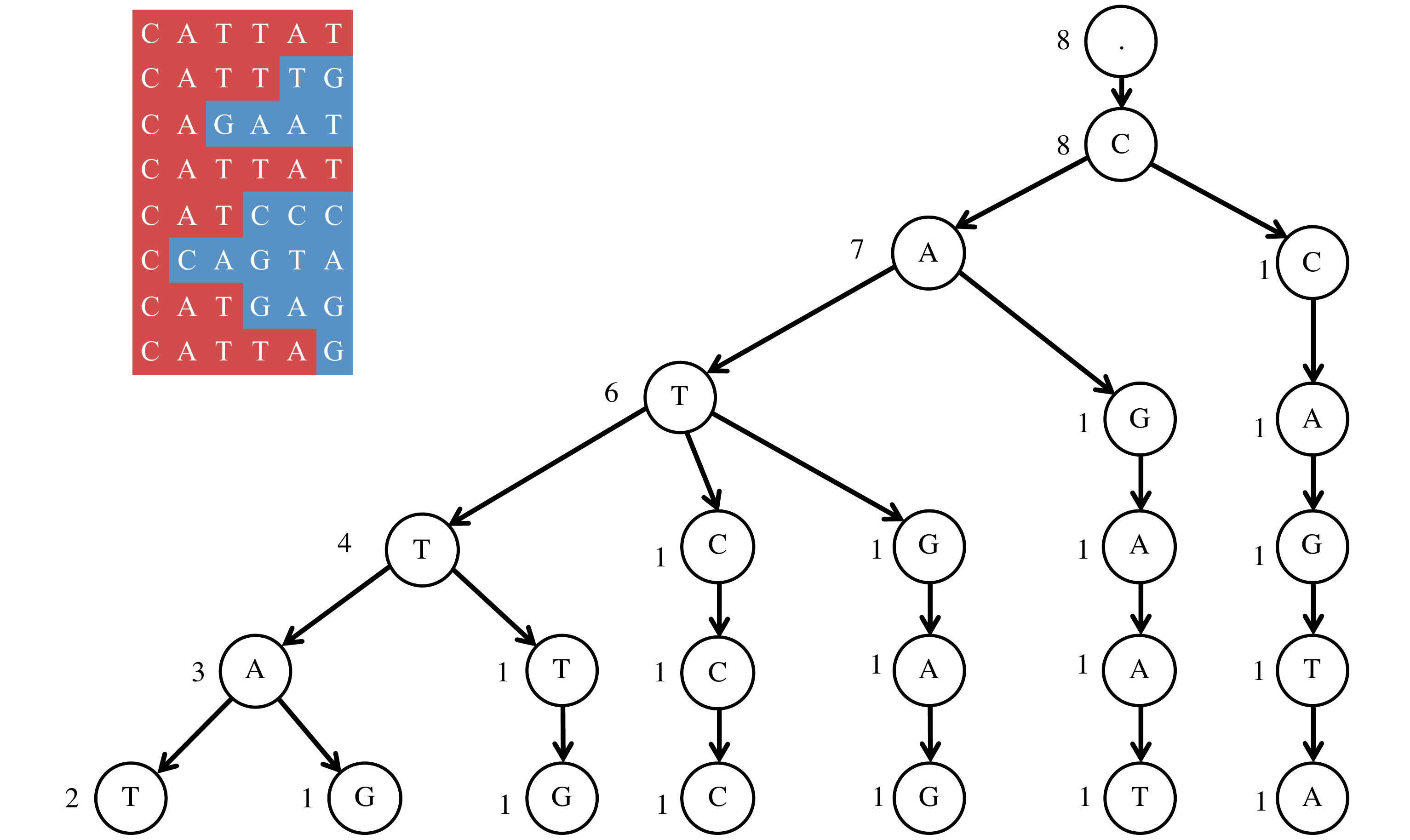}
\caption{Illustration of the algorithm for solving the String Reconstruction Problem in the \emph{TrimSuffixAndExtend} model. The set of traces is shown on the left, and their trie is shown on the right. The string associated with each vertex is the one that is formed by traversing the trie from the root node to the vertex. The values of $sim_t(C,s)$ for all vertices are shown.}
\label{fig_exact_algo}
\end{figure*}

\newtext{
For all strings in $C$ and all values of $t$, the quantities $sim_t(C,s)$ can be computed during the construction of the trie as follows. Traces are added sequentially to construct the trie. In addition to $t$-prefixes, each vertex also stores $sim_t(C,s)$ which is initialized to $1$ for a new vertex. For example, in Figure \ref{fig_exact_algo}, we start with an empty trie and first add the trace "CATTAT" by creating six new vertices, each representing one of the six $t$-prefixes. At this point, the trie contains only one string, and for all vertices, we have $sim_t(C,s) = 1$. Then, we add the next trace ``CATTTG''. For $t\leq4$, the $t$-prefixes of ``CATTAT" and ``CATTTG'' coincide. In other words, they share the first four vertices in the trie. For all vertices that are traversed while inserting a new trace, the values of $sim_t(C,s)$ are updated by adding $1$ to the current values. For new vertices, like before, the values of $sim_t(C,s)$ are initialized to~$1$. In this example, for the vertices representing $t$-prefixes ``C'', ``CA'', ``CAT'', and ``CATT'', the value of $sim_t(C,s)$ will be updated to $2$, whereas for the two new vertices representing $t$-prefixes ``CATTT'' and ``CATTTG'', the values of $sim_t(C,s)$ will be $1$. All traces are inserted to the trie in this manner. We can thus compute all $sim_t(C,s)$ values during the construction of the trie with complexity $O(|s| \cdot T)$. 
After the construction of the trie, all quantities $\text{score}(C^t \mid s^t)$ can then be computed by a single Depth-First Search using Eq. (\ref{eq_recurion}). 
}

\subsubsection*{\textbf{TrimSuffixAndExtend model with multiple seeds}} 
Next, we consider a modified \textit{TrimSuffixAndExtend} model with a seed-set $S = \{s_1, s_2, \ldots, s_M\}$. Traces are generated via a two step approach. First, a string $s_i \in S$ is chosen uniformly randomly from $S$. Then, $s_i$ is modified to generate a trace $c$ according to the \textit{TrimSuffixAndExtend} model. We note that $S$ can either be an arbitrary set of $M$ strings (worst-case) or the strings in $S$ can be chosen independently and uniformly from the universe of possible strings (average-case). 
Note that the above model is described for a uniform distribution over the seed strings. 
 In the real VDJ recombination process, various D genes contribute to immunoglobulin genes with varying propensities. To incorporate this fact, the above model can be reformulated by considering an arbitrary distribution on the seed strings. 
\begin{quote}
\vspace{.1in}
\textbf{Trace Reconstruction with Multiple Seeds Problem in the \textit{TrimSuffixAndExtend} model}\\
\textbf{Input:} A trace-set $C$ generated from an unknown set of $M$ seed strings of the same length according to the \textit{TrimSuffixAndExtend} model.  \\
\textbf{Output:} A set of strings $S = \{s_1, s_2, \ldots , s_M\}$ maximizing $\pCS$.
\vspace{.1in}
\end{quote}

\subsubsection*{\textbf{The MINING-D heuristic algorithm}}
Although the trace reconstruction problem can be efficiently solved in the \textit{TrimSuffixAndExtend} model, it is unclear how to generalize the algorithm for the more complex models with multiple D genes and varying lengths of modified strings. Bhardwaj et. al.~\cite{bhardwaj2019mining_d} propose a practical greedy heuristic for this model that, while being suboptimal, motivates practical algorithms for more complex models. 

For the \textit{TrimSuffixAndExtend} model, the algorithm starts with an empty string and at step $j$  extends it on the right by the most abundant symbol in $C$  at position $j$  and discards from $C$  the strings that have symbols that are not the most abundant symbols at position $j$. This procedure repeats until the length of the resulting string equals the length of the seed string $s$. This greedy algorithm, however, cannot be directly used in practice because  (a) the CDR3s are formed by multiple D genes, (b) the number of D genes is unknown {\em a priori}, (c) the D genes have different lengths that are unknown, (d) CDR3s generated by the same D gene can have different lengths. 

The MINING-D algorithm \cite{bhardwaj2019mining_d}, inspired by the above greedy algorithm, considers the complexities of the real immunogenomics data. 
It uses the observation that, although D genes typically get truncated on both sides during the VDJ recombination process, their truncated substrings are often present in the newly recombined genes, and, hence, the CDR3s. Therefore, we expect the truncated substrings of D genes to be highly abundant in a CDR3 dataset. MINING-D starts by finding the most abundant $k$-mers (a $k$-mer is a string of length $k$). It then extends them on both sides using the greedy algorithm to recover entire D genes that contain highly abundant $k$-mers as substrings. MINING-D defines a probabilistic stopping rule as the lengths of the D genes are not known {\em a priori}. This stopping rule also allows us to recover D genes of different lengths. Since some abundant $k$-mers can be substrings of multiple D genes, MINING-D allows multiple extensions from each $k$-mer in the extension procedure. 

We next introduce models that incorporate more complexities of the VDJ recombination process, leading up to the model that mimics the real formation of an immunoglobulin gene from a set of V, D, and J genes. To the best of our knowledge, these models have not been studied in the literature and brute-force search is the only known exact solution to trace reconstruction in these models.

\subsection{Toward a biologically adequate model for D gene reconstruction}
\label{sec_immunology_adequate}

\subsubsection*{\textbf{$\text{SuffixExtend}_t$(TrimSuffix)}}

Unlike the \emph{TrimSuffixAndExtend} model, the \emph{$\text{SuffixExtend}_t$(TrimSuffix(s))} model (Figure \ref{fig_d_models}b) generates traces of varying lengths from a single seed string $s$. Let $s_{trim}$ be the substring of $s$ that remains after the operation \emph{TrimSuffix} is applied on $s$.
Then, $\pcs$ is given by
\begin{align*}
\pcs &= \sum_{k=0}^{|s|} \Pr(c,|s_{trim}|=k \mid s) \\
&= \sum_{k=0}^{|s|} \Pr(|s_{trim}|=k \mid  s) \Pr(c \mid  s, |s_{trim}|=k) \\
&=  \frac{1}{(|s|+1)}  \sum_{k=0}^{|s|}  \Pr(c \mid  s, |s_{trim}|=k)
\end{align*}
Let $m$ be the length of the longest shared prefix between $c$ and $s$, as before. Then,  $\Pr(c \mid  s, |s_{trim}|=k)$  is non-zero only if $ |c|-t \leq k \leq m$ and can be written as 
\begin{equation*}
\Pr(c \mid  s, |s_{trim}|=k) =
   \begin{cases}
     \frac{1}{(t+1)|\caliA|^{|c|-k}} &\text{if  $ |c|-t \leq k \leq m$}  \\
     0 &  \text{otherwise}
   \end{cases}
\end{equation*}
Thus $\pcs$ is zero if $m < |c|-t$. Otherwise, 
\begin{equation}
\pcs = \frac{1}{(|s|+1)(t+1)} \sum_{k=  	(|c|-t)^+}^m
\frac{1}{ {|\caliA|}^{|c| -k}} \label{eq_suffix_extend_trim}
\end{equation}
where $x^+ = \max(x,0)$.

\begin{quote}
\vspace{.1in}
\textbf{Trace Reconstruction Problem in the \emph{$\text{SuffixExtend}_t$(TrimSuffix(s))} model}\\
\textbf{Input:} A trace-set $C$ generated from an unknown seed string according to the \emph{$\text{SuffixExtend}_t$(TrimSuffix(s))} model.\\
\textbf{Output:} A string maximizing $\mathbf{Pr}(C\mid s)$.
\vspace{.1in}
\end{quote}

\subsubsection*{\textbf{$\text{SuffixExtend}_t$($\text{Mutate}_{\epsilon}$(TrimSuffix))}}
We now consider a slightly more realistic model for trace generation that incorporates somatic hypermutations (Figure \ref{fig_d_models}c). The probability  $\pcs$ that a seed string $s$ generates a trace $c$ is given by 

\begin{IEEEeqnarray*}{rCl}
\pcs & = & \frac{1}{(|s|+1)(t+1)} \\ 
&& \times\> \sum_{k=  (|c|-t)^+}^{|s|}  \frac{(1-\epsilon)^{k - d_k} {(\epsilon/(|\caliA|-1))}^{d_k}}  { {|\caliA|}^{|c| -k}}
\end{IEEEeqnarray*}

where $d_k$ is the Hamming distance between the prefixes of $c$ and $s$ of length $k$. 

\begin{quote}
\vspace{.1in}
\textbf{Trace Reconstruction Problem in the \emph{$\text{SuffixExtend}_t$($\text{Mutate}_{\epsilon}$(TrimSuffix))} model}\\
\textbf{Input:} A trace-set $C$ generated from an unknown seed string according to the \emph{$\text{SuffixExtend}_t$($\text{Mutate}_{\epsilon}$(TrimSuffix)) } model.\\
\textbf{Output:} A string maximizing $\mathbf{Pr}(C\mid s)$.
\vspace{.1in}
\end{quote}


\subsubsection*{\textbf{TrimAndExtend}} 
In all the models above, only the suffix of the seed string gets trimmed in the first step. In contrast, during the VDJ recombination process, the D gene gets trimmed from both sides. We will thus consider the \textit{TrimAndExtend} model  (Figure \ref{fig_d_models}d) for generating a trace $c$ from a seed string $s$. 

Since strings $s$ and $c$ have the same length, their comparison results in a binary comparison vector where 1s (0s) correspond to the match (mismatch) positions. Let $t(i)$ denote the length of the continuous run of 1s starting at position $i+1$ in the comparison vector. The probability that a seed string $s$ generates a trace $c$ is given by 
\begin{equation*}
\pcs = \frac{2}{(|s|+1)(|s|+2)} \sum_{i=0}^{|s|} 
\left(
\sum_{k= |s|-i-t(i)}^{|s|-i} \frac{1}{|\caliA|^{i+k}}
\right)
\end{equation*}
\begin{quote}
\vspace{.1in}
\textbf{Trace Reconstruction Problem in the \textit{TrimAndExtend} model}\\
\textbf{Input:} A trace-set $C$ generated from an unknown seed string according to the \emph{TrimAndExtend} model.\\
\textbf{Output:} A string maximizing $\mathbf{Pr}(C\mid s)$.
\vspace{.1in}
\end{quote}

\subsubsection*{\textbf{$\text{Mutate}_{\epsilon}$(TrimAndExtend)}} 
We now consider a model that incorporates mutations in the \textit{TrimAndExtend} model (Figure \ref{fig_d_models}e). 
Let $\text{substring}_{l,k}(s)$ be the substring of the seed string $s$ where the prefix of length $l$ and the suffix of length $k$ have been trimmed. 
The probability that a seed string $s$ generates a trace $c$ in the \textit{$\text{Mutate}_{\epsilon}$(TrimAndExtend)} model is given by 


\begin{IEEEeqnarray*}{rCl}
\pcs & = & \frac{2}{(|s|+1)(|s|+2)}  \times \\ 
&& \> \sum_{i=0}^{|s|} 
\left(
\sum_{k= 0}^{|s|-i} \frac{ {(\epsilon/(|\caliA|-1))}^{d_{i,k}}   {(1-\epsilon)}^{|s| -i -k -d_{i, k}  }}  {|\caliA|^{i+k}}
\right)
\end{IEEEeqnarray*}
where $d_{l,k}$ is the Hamming distance between $\text{substring}_{l,k}(c)$ and $\text{substring}_{l,k}(s)$. 
\begin{quote}
\vspace{.1in}
\textbf{Trace Reconstruction Problem in the \textit{$\text{Mutate}_{\epsilon}$(TrimAndExtend)} model}\\
\textbf{Input:} A trace-set $C$ generated from an unknown seed string according to the \emph{$\text{Mutate}_{\epsilon}$(TrimAndExtend)} model.\\
\textbf{Output:} A string maximizing $\mathbf{Pr}(C\mid s)$.
\vspace{.1in}
\end{quote}

\subsubsection*{\textbf{$\text{Extend}_t(\text{Mutate}_{\epsilon}$(Trim))}} 
The biologically adequate model for generating traces from a seed string is the \textit{$\text{Extend}_t(\text{Mutate}_{\epsilon}$(Trim))} model illustrated in Figure \ref{fig_trim_mutate_extend_model}. 
This model is more complex than the previous ones as it requires consideration of all possible pairs of equally sized substrings of the seed string and the trace. 
Note that in all previous models, the traces either had the same length as the seed string, or were aligned with the seed string on the left. 
Let  $\text{sub}^l(s)$ denote all the substrings of $s$ of length~$l$ and 
 $\text{sub}^l_t(c)$ denote all substrings of $c$ of length $l$ such that the number of symbols in $c$ before or after the substring do not exceed $t$. 
Then, the probability that a seed string $s$ generates a trace $c$ in the \textit{$\text{Extend}_t(\text{Mutate}_{\epsilon}$(Trim))} model is given by 


\begin{IEEEeqnarray}{rCl}
\pcs & = & \frac{1}{(t+1)^2} \frac{2}{(|s|+1)(|s|+2)} \nonumber \\ 
&& \times\> \sum_{l=0}^{\min{(|s|, |c|)}} \frac{1}{|\caliA|^{|c| -l}} \nonumber \\ 
&& \left(
\sum_{\substack{\bar{s} \in \text{sub}^l(s)\\ \bar{c} \in \text{sub}^l_t(c)}}
(1 - \epsilon)^{l - d_{\bar{s}, \bar{c}}}
{\left(\frac{\epsilon}{|\caliA|-1}\right)}  ^{d_{\bar{s}, \bar{c}}} 
\right),
\label{eq_extend_mutate_trim}
\end{IEEEeqnarray}
where $d_{s_1, s_2}$ is the Hamming distance between strings $s_1$ and~$s_2$. 

\begin{quote}
\vspace{.1in}
\textbf{Trace Reconstruction Problem in the \textit{$\text{Extend}_t(\text{Mutate}_{\epsilon}$(Trim))} model}\\
\textbf{Input:} A trace-set $C$ generated from an unknown seed string according to the \emph{$\text{Extend}_t(\text{Mutate}_{\epsilon}$(Trim))} model.\\
\textbf{Output:} A string maximizing $\mathbf{Pr}(C\mid s)$.
\vspace{.1in}
\end{quote}

\subsection{Trace Reconstruction of V, D, and J genes}
\label{sec_immunology_vdj_reconstruction}
\begin{figure*}[ht]
     \centering
     \begin{subfigure}[b]{0.7\textwidth}
         \includegraphics[width=\textwidth]{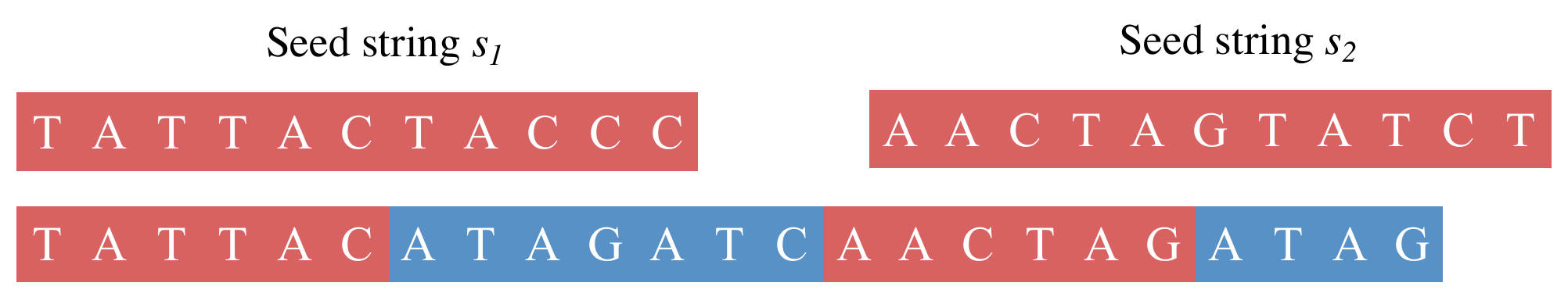}
         \caption{\textit{$\text{SuffixExtend}_t$(TrimSuffix($s_1$))*$\text{SuffixExtend}_t$(TrimSuffix($s_2$))}}
         \label{fig_vdj_1}
     \end{subfigure}
     \medskip \vspace{.25in}
     \begin{subfigure}[b]{0.7\textwidth}
         \includegraphics[width=\textwidth]{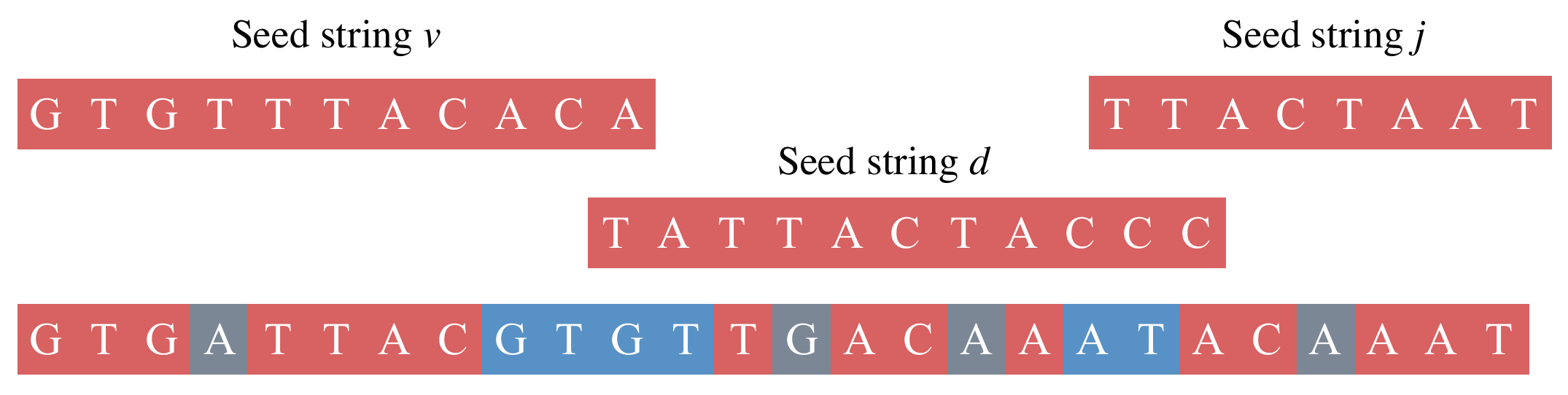}
         \caption{\textit{$\text{Mutate}_{\epsilon}$(TrimSuffix(v)*$\text{Extend}_t$(Trim(d))* TrimPrefix(j))}}
         \label{fig_vdj_2}
     \end{subfigure}
     
        \caption{Trace generation that involves concatenation of multiple seed strings. Insertions are shown in light blue, hypermutations are shown in green. The most general model for the VDJ recombination is shown in (b). }
        \label{fig_vdj_models}
\end{figure*}

Above, we considered the trace reconstruction problems that are relevant to generating a CDR3 from a D gene. We will now consider more complex trace reconstruction problems that model concatenation of V, D, and J genes to form an entire immunoglobulin gene (Figure \ref{fig_vdj_models}). We will start from the simplest problem when each trace represents a concatenation of just two traces generated by two different seed strings.


\subsubsection*{\textbf{$\text{SuffixExtend}_t$(TrimSuffix)*$\text{SuffixExtend}_t$(TrimSuffix)}}
We first consider a model when two seed strings $s_1, s_2$ of equal length $n$ generate a single trace $c$ according to the \textit{$\text{SuffixExtend}_t$(TrimSuffix($s_1$))*$\text{SuffixExtend}_t$(TrimSuffix($s_2$))} model (Figure \ref{fig_vdj_1}). 
Let $\text{prefix}_l(s)$ and $\text{suffix}_l(s)$ be the prefix and suffix of string $s$ of length $l$. The probability  that the seed strings $s_1$ and $s_2$ generate a trace $c$ is given by 
\begin{IEEEeqnarray}{rCl}
\mathbf{Pr}(c \mid s_1, s_2) & =  \sum_{l=0}^{|c|} & 
\mathbf{Pr}(\text{prefix}_l(c) \mid s_1)  \times \nonumber \\ 
&&   \mathbf{Pr}(\text{suffix}_{|c|-l}(c) \mid s_2) 
\label{eq_concat_suffexttrim}
\end{IEEEeqnarray}
where $\mathbf{Pr}(\text{prefix}_l(c) \mid s_1) $ is defined according to the \textit{$\text{SuffixExtend}_t$(TrimSuffix)} model (Eq. \ref{eq_suffix_extend_trim}) if $l \leq n+t$ and $0$ otherwise.  
$\mathbf{Pr}(\text{suffix}_{|c| -l}(c) \mid s_2)$ is defined similarly. 

\begin{quote}
\vspace{.1in}
\textbf{Trace Reconstruction Problem in the \textit{$\text{SuffixExtend}_t$(TrimSuffix)*$\text{SuffixExtend}_t$(TrimSuffix)} model}\\
\textbf{Input:} A trace-set $C$ generated from two unknown seed strings according to the  \emph{$\text{SuffixExtend}_t$(TrimSuffix)*$\text{SuffixExtend}_t$(TrimSuffix)} model.\\
\textbf{Output:} Strings $s_1$ and $s_2$  maximizing $\mathbf{Pr}(C\mid s_1, s_2)$.
\vspace{.1in}
\end{quote}

\subsubsection*{\textbf{$\text{SuffixExtend}_t$(TrimSuffix)*$\text{SuffixExtend}_t$(TrimSuffix) model with multiple seeds}}
Next, we consider a modification of the above model where each trace is generated by two sets of seed strings of the same length $n$, $S_1 = \{ s_1^1, s_1^2, \ldots, s_1^{M_1}\} $ and
$S_2 = \{ s_2^1, s_2^2, \ldots, s_2^{M_2}\} $, rather than a pair of seed strings. 
 Seed strings $s_1$ and $s_2$ are  randomly chosen (from the sets $S_1$ and $S_2$ according to a uniform distribution) and the chosen strings generate a trace according to the  \textit{$\text{SuffixExtend}_t$(TrimSuffix)*$\text{SuffixExtend}_t$(TrimSuffix)} model. 

\begin{quote}
\vspace{.1in}
\textbf{Trace Reconstruction with Multiple Seeds Problem in the   \textit{$\text{SuffixExtend}_t$(TrimSuffix)* $\text{SuffixExtend}_t$(TrimSuffix)} model}\\
\textbf{Input:} A trace-set $C$ generated from two unknown sets containing $M_1$ and $M_2$ seed strings according to the   \emph{$\text{SuffixExtend}_t$(TrimSuffix)* $\text{SuffixExtend}_t$(TrimSuffix) } model.\\
\textbf{Output:} A set of $M_1$ seed strings and a set of $M_2$ seed strings maximizing $\mathbf{Pr}(C\mid S_1, S_2)$.
\vspace{.1in}
\end{quote}

\subsubsection*{\textbf{VDJ recombination model (single v, d, and j seed strings)}} 
We now consider a model when three strings $v$, $d$, and~$j$ of length $n_v$, $n_d$, and $n_j$ respectively generate a trace $c$ according to the \textit{$\text{Mutate}_{\epsilon}$(TrimSuffix(v)*$\text{Extend}_t$(Trim(d))* TrimPrefix(j))} model (Figure \ref{fig_vdj_2}). 
Here, \textit{TrimPrefix(s)} is defined similarly to \textit{TrimSuffix(s)}, where an integer $k$ is sampled uniformly from $[0, |s|]$, and the prefix of $s$ of length $k$ is trimmed. 
However, like the \textit{$\text{Extend}_t(\text{Mutate}_{\epsilon}$(Trim)))} model, it is a complicated model because one must consider all triples of substrings of the trace $c$. 
The probability $\mathbf{Pr}(c \mid v, d, j)$  that the seed strings $v$, $d$, and $j$ generate a trace c is given by 


\begin{IEEEeqnarray*}{rCl}
\mathbf{Pr}(c \mid v, d, j) & =  \sum_{i=0}^{n_v} \sum_{k=0}^{\min(|c|-i, n_j)} 
&  P_1 (\text{prefix}_i(c) \mid v)  \times \\ 
&&  P_2 (\text{substring}_{i, k}(c) \mid d) \times  \\
&&  P_3 (\text{suffix}_k(c) \mid j)  ,
\end{IEEEeqnarray*}
where $P_1 (\text{prefix}_i(c) \mid v)$ is given by 
\begin{equation*}
P_1 (\text{prefix}_i(c) \mid v) = \frac{1}{n_v + 1} {(\epsilon/(|\caliA|-1))}^{d_i} {(1-\epsilon)}^{i - d_i}
\end{equation*}
where $d_i$ is the Hamming distance between $\text{prefix}_i(c) $ and $\text{prefix}_i(v)$. $P_2 (\text{substring}_{i, j}(c) \mid d)$ is defined as in Eq.~(\ref{eq_extend_mutate_trim}).  
$P_3$ is defined similarly to $P_1$.  

\begin{quote}
\vspace{.1in}
\textbf{Trace Reconstruction Problem in the \textit{VDJ recombination} (single \textit{v}, \textit{d}, and \textit{j} seed strings) model}\\
\textbf{Input:} A trace-set $C$ generated from three unknown seed strings according to the \emph{VDJ recombination } model.\\
\textbf{Output:} Three strings $s_1$ , $s_2$, and $s_3$  maximizing $\mathbf{Pr}(C\mid v, d, j)$.
\vspace{.1in}
\end{quote}

\subsubsection*{\textbf{VDJ recombination model (multiple v, d, and j seed strings)}} 
We will now consider a model when three seed-sets $V = \{v_1, v_2, \ldots, v_{M_v}\}$, $D = \{ d_1, d_2, \ldots, d_{M_d} \}$, and $J = \{ j_1, j_2, \ldots , j_{M_j} \}$ generate a trace $c$ according to the following model. One string from each of the sets $V$, $D$, and $J$ is uniformly randomly chosen and the chosen strings $v$, $d$, and $j$ generate a trace according to the VDJ recombination model. The probability  that a trace $c$ is generated by seed strings in $V$, $D$, and $J$ is given by 
\begin{equation*}
\mathbf{Pr}(c \mid V, D, J) = \frac{1}{M_v M_d M_j} 
\sum_{v \in V} \sum_{d \in D} \sum_{j \in J} \mathbf{Pr}(c \mid v, d, j) 
\end{equation*}
\begin{quote}
\vspace{.1in}
\textbf{Trace Reconstruction Problem in the \textit{VDJ recombination} (multiple \textit{v}, \textit{d}, and \textit{j} seed strings) model}\\
\textbf{Input:} A trace-set $C$ generated from three unknown seed-sets (containing $M_v$, $M_d$, and $M_j$ strings respectively) according to the \emph{VDJ recombination } model.\\
\textbf{Output:} Set $S_1$ with $M_v$ strings, set $S_2$ with $M_d$ strings, and set $S_3$ with $M_j$ strings maximizing $\mathbf{Pr}(C\mid S_1,S_2, S_3).$
\end{quote}

\section{Trace Reconstruction problems for DNA Data storage}
\label{sec_dna_storage}

\begin{figure}[t]
\centering
\includegraphics[width = \columnwidth]{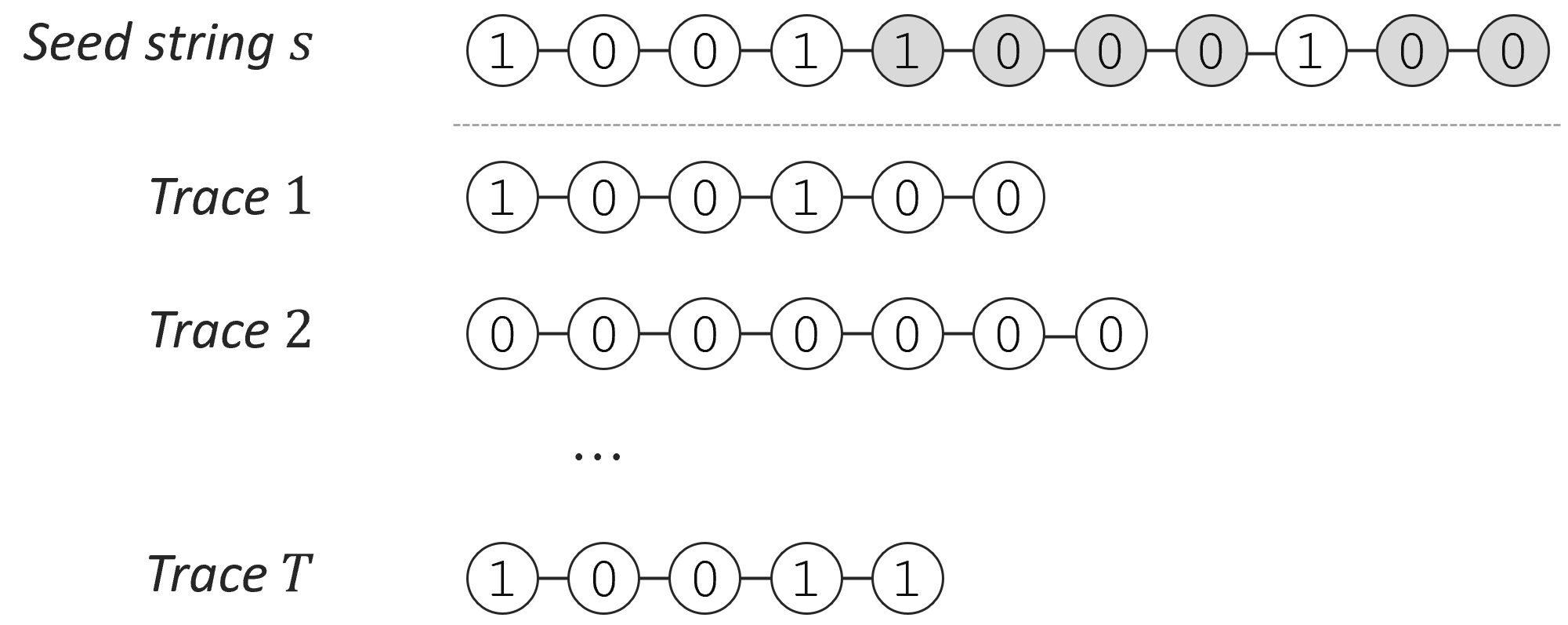}
\caption{Seed string and example traces from the deletion channel. Gray circles indicate the deleted bits to generate the bottom trace.}
\label{deletion-channel}
\end{figure}

A popular formulation of trace reconstruction considers the {\em deletion channel}, where random symbols in the seed string $s$ are deleted independently with probability $q$ and $0 < q < 1$ is the {\em deletion probability}. This produces a trace~$c$ representing a random subsequence of~$s$. This process is repeated independently $T$ times to produce a random trace-set $C$ (Figure~\ref{deletion-channel}). The trace reconstruction algorithm takes the traces (without any information about which symbols were deleted from the seed string), the length of the seed string, and the deletion probability as an input. For simplicity, we focus on binary seed strings, while the definitions can be extended to larger alphabets.

The maximum likelihood solution would output the string $s$ that maximizes $\mathbf{Pr}(C \mid s)$ for the given trace-set $C$. We first consider the probability $\mathbf{Pr}(c \mid s)$ for a single trace~$c$. Let $N_s(c)$ denote the number of times $c$ appears as a subsequence of $s$.  For example, if $s = 11010$ then $c = 110$ appears $N_s(c) = 4$ times, corresponding to the subsequences $\{110\bullet\bullet,\ 11\bullet\bullet~0,\ 1\bullet\bullet~10,\ \bullet~1\bullet\vspace{-.1ex}10\}$, where ‘$\bullet$’ denotes a deleted symbol. The value of  $N_s(c)$ can be computed using dynamic programming~\cite{compeau2015bioinformatics}. Recalling that $|s|$ denotes the length of a string, the probability $\mathbf{Pr}(c \mid s)$ can be computed as follows 
$$
  \mathbf{Pr}(c \mid s) = N_s(c) \cdot q^{|s|-|c|} (1-q)^{|c|}.
$$
Since each trace in $C$ is produced independently, we have that 
$$
  \mathbf{Pr}(C \mid s) = \prod_{c \in C} \mathbf{Pr}(c \mid s).
$$
The value $\mathbf{Pr}(C \mid s)$ can be calculated for any fixed $s$. However, the optimization problem that determines the optimal $s$ is challenging. Designing an efficient algorithm (with time polynomial in $|C|$ and $|s|$) that outputs a string~$s$ maximizing $\mathbf{Pr}(C \mid s)$ is an open question. Partial results are known when $|C|$ is very small~\cite{mitzenmacher-survey, sabary2020error, srinivasavaradhan2020algorithms,srinivasavaradhan2019symbolwise, srinivasavaradhan2018maximum}. 

We focus on the success probability in this section, and we also restrict to length $n$ seed strings. We define the {worst-case success probability} of an algorithm $A$ over all binary strings of length $n$ as
 $$
   P_A(n, T)  = \min_s P_A(s, T).
 $$
Similarly, the {average-case success probability} of $A$ over all binary strings of length $n$ is
 $$
   \widetilde P_A(n, T)  = \frac{1}{2^n} \cdot \sum_s P_A(s, T).
 $$

\subsubsection*{\textbf{Trace Complexity}}
Most previous work provides information-theoretic results in terms of the {\em trace complexity}, which is the minimum value of $T$ such that there exists an algorithm with success probability at least the given \textit{ReconstructionRate}. This will depend on the deletion probability $q$. For any fixed \textit{ReconstructionRate}, the number of input traces must be at least the trace complexity for the algorithmic problem to be feasible. It is often convenient to fix the \textit{ReconstructionRate} to a default value, such as \textit{ReconstructionRate} $= 0.95$. This does not affect the trace complexity too much because arbitrarily large \textit{ReconstructionRate} can be achieved by increasing the number of traces by a logarithmic factor (taking a majority vote over several trials). Therefore, we define the {\em worst-case trace complexity} as
$$
 T_q(n) = \argmin\left\{ T\ \bigm\vert\ \max_A  P_A(n, T) \geq 0.95 \right\}
$$
and the {\em average-case trace complexity} as
$$
  \widetilde T_q(n) = \argmin \left\{T\ \bigm\vert\ \max_A \widetilde P_A(n, T) \geq 0.95 \right\}.
$$

The trace complexity may depend on the error rate. Certain algorithms only succeed when the deletion probability decreases as a function of the length $n$ of the seed string. 
Historically, the initial results assume that the deletion probability scales inversely with $n$, e.g., $q =  O(1/\sqrt{n})$ or $q = O(1/\log n)$~\cite{BatuKannan04-RandomCase, HolensteinMPW08, viswanathan}. These results have been later strengthened to handle a constant rate of deletions, e.g., $q=0.5$~\cite{de2019optimal,nazarov2017trace,holden2018subpolynomial}. The extent to which the deletion probability impacts the trace complexity remains unknown in general.

For simplicity, we restrict our attention to the deletion channel, but many of the results that we discuss also extend to a more general error model that includes insertions and substitutions~\cite{de2019optimal, holden2018subpolynomial, nazarov2017trace, viswanathan}. We refer the reader to the following surveys for other error models and related theoretical open questions~\cite{cheraghchi2020overview, mitzenmacher-survey}.

\section{Theoretical Results on Trace Reconstruction}
\label{sec_theoretical_results}

We survey theoretical results for reconstructing a seed string $s$ of length $n$. We begin with three variants depending on the nature of the unknown string: it can be arbitrary (worst-case); it can be chosen uniformly at random (average-case); or, it can be chosen from a predefined set of encoded strings (\textit{coded trace reconstruction}). For each variant, we first present a formal problem statement. The information-theoretic goal is to determine the values of the parameters $T,q,n,$ and \textit{ReconstructionRate} for which the problem is solvable. The next step is to design an efficient algorithm for such cases. In the latter half of this section, we also mention generalizations to multiple strings and to higher-order structures (such as trees). We conclude with a brief description of some recent practical developments. Throughout, we use $\hat s =A(C)$ to abbreviate the output of a reconstruction algorithm $A$ on an input trace-set $C$. 

\subsubsection*{\textbf{Worst-case trace reconstruction}}
We first describe the case where the seed string $s$ is arbitrary, and the success probability is calculated over the randomness in generating the trace-set $C$. 
\begin{quote}
\vspace{.1in}
\textbf{Worst-Case Trace Reconstruction Problem for the Deletion Channel}\\
\textbf{Input:} A random trace-set $C$ of size $T$ generated from a seed string $s$ of length $n$ according to the deletion channel model with deletion probability $q$, as well as the \textit{ReconstructionRate}.\\
\textbf{Output:} A string $\hat s$ such that $\hat s = s$ with success probability at least \textit{ReconstructionRate}.
\vspace{.1in}
\end{quote}

\newtext{
The current best trace complexity for worst-case strings is $T_q(n) = \mathrm{exp}(O(n^{1/5}\log^5 n))$ when the deletion probability $q$ is at most $1/2$~\cite{chase2020new}. When $q \in (1/2,1)$, then the known result is $T_q(n) = \mathrm{exp}(O(n^{1/3}))$~\cite{de2019optimal, nazarov2017trace}. The latter result uses a {\em mean-based algorithm} that first pads each trace with trailing zeros so that the length equals the seed length $n$ (here, we consider a binary alphabet). Then, the mean of the traces is computed by summing the padded traces coordinate-wise and normalizing by the number of traces (i.e., this computes the fraction of ones in each position). It is known that when the number of traces is at least $\mathrm{exp}(O(n^{1/3}))$ then these means suffice to determine the unknown string with high success probability~\cite{de2019optimal, nazarov2017trace}. The improvement to $T_q(n) = \mathrm{exp}(O(n^{1/5}\log^5 n))$ when $q\leq 1/2$ uses a similar algorithm, with the subtle difference and important difference that certain substring frequencies are approximated instead of single bits~\cite{chase2020new}.}


An intriguing aspect of the worst-case result is the use of techniques from complex analysis. The elegant argument involves expressing the mean-based statistics (from averaging the padded traces) in terms of a complex-valued generating function (whose coefficients are determined by the seed string and deletion probability). The aim is to lower bound the statistical distance between trace-sets that are generated from distinct seed strings. It is fairly easy to show that the maximum modulus of this function in a certain arc of the complex unit disk provides such a lower bound. Then, to complete the proof, the authors use prior results on Littlewood polynomials~\cite{borwein1997littlewood, borwein1999littlewood}.
This argument serves as the basis of a trace reconstruction algorithm with running time proportional to the number of traces.
Surprisingly, the bound is tight for mean-based algorithms, in the sense that $\mathrm{exp}(\Omega(n^{1/3}))$ traces are necessary if an algorithm uses only the coordinate-wise means~\cite{de2019optimal, nazarov2017trace}.
These results have further inspired the use of related generating functions to derive improved bounds for other statistical learning problems~\cite{krishnamurthy2020algebraic, krishnamurthy2019sample}.

Improvements to the trace complexity are known for a very small deletion probability; if each bit is deleted with probability less than $n^{-1/2-\delta}$ for a small constant $\delta$, then a nearly-linear number of traces suffice~\cite{HolensteinMPW08}. We note that mean-based algorithms extend to handle insertions and substitutions as well~\cite{de2019optimal, nazarov2017trace}. It is an open question to determine the smallest deletion probability such that a polynomial number of traces suffice.
When the deletion probability does not decrease with $n$ (e.g., $q=0.5$), then lower bounds on the trace complexity are known. Previous work shows $T_{0.5}(n) = \widetilde \Omega(n^{3/2})$ traces are necessary~\cite{Chase19, holden2020lower}, where the $\widetilde \Omega$ notation hides polylog factors. 

The central open problem is to close the exponential gap between upper and lower bounds on the worst-case trace complexity. A first step could be to better understand which seeds strings are the most challenging to reconstruct. For many algorithms, simple strings demonstrate that the current analysis is tight. However, other methods readily reconstruct these strings. For example, the $\widetilde \Omega(n^{3/2})$ lower bound is derived for the task of distinguishing a pair of alternating strings with two flipped bits, e.g.,
\begin{align*}
  1010\cdots10\underline{10}10\cdots1010 \\
  1010\cdots10\underline{01}10\cdots1010
\end{align*}
Telling apart these strings using traces is straightforward, and an algorithm using $\widetilde O(n^{3/2})$ traces is known. Hence, the lower bound for this pair is nearly tight~\cite{Chase19, holden2020lower}. Another futile attempt comes from considering a uniformly random string. In many areas, the probabilistic method suffices to identify difficult instances~\cite{alon2004probabilistic, arora2009computational}. For reconstruction problems, the opposite is often true: random objects can be reconstructed with less information than worst-case instances~\cite{bollobas1990almost, przykucki2019shotgun, radcliffe1998reconstructing}. In particular, random strings are easier to reconstruct, as we will now see.

\subsubsection*{\textbf{Average-case trace reconstruction}}
We move on to consider the case when the seed string $s$ is a uniformly random length $n$ string. In this case, the seed string is chosen randomly before generating each set of traces, and the success probability is calculated with respect to both the trace-set generation and the choosing of the seed string. 
\begin{quote}
\vspace{.1in}
\textbf{Average-Case Trace Reconstruction Problem for the Deletion Channel}\\
\textbf{Input:} A random trace-set $C$ of size $T$ generated from a uniformly random seed string $s$ of length $n$ according the deletion channel model with deletion probability $q$, as well as the \textit{ReconstructionRate}.\\
\textbf{Output:} A string $\hat s$ such that $\hat s = s$ with success probability at least \textit{ReconstructionRate}.
\vspace{.1in}
\end{quote}

The current best upper bound on the trace complexity is $\widetilde T_q(n) = \mathrm{exp}(O(\log^{1/3}n))$ for uniformly random strings, and this holds for any deletion probability $q$ bounded away from one~\cite{holden2018subpolynomial}. This upper bound is exponentially better than the result for worst-case strings~\cite{de2019optimal, nazarov2017trace}. The lower bound for average-case reconstruction shows that $\widetilde T_{0.5}(n) = \widetilde \Omega(\log^{3/2} n)$ traces are necessary to reconstruct a random string with constant deletion probability, where here the $\widetilde \Omega$ notation hides $\log \log n$ factors~\cite{Chase19, holden2020lower}. When the deletion probability scales inverse-logarithmically with $n$, then logarithmic upper bounds on the average-case trace complexity are known~\cite{BatuKannan04-RandomCase, viswanathan}.

The algorithms for average-case reconstruction are much more involved than the current methods for worst-case reconstruction. Instead of relying only on statistical quantities, the algorithm iteratively reconstructs the seed string one character at a time. At the beginning, a small number of traces are used to learn a short prefix exactly. 
This partial reconstruction then serves as an anchoring method to approximately align the traces. When the seed string is random, its short substrings are locally unique with high probability, and therefore, such alignments can be reliable. The algorithm moves left-to-right and employs a worst-case algorithm to reconstruct the next bit. 
This general approach, along with a careful analysis of the alignment process, led to an algorithm that requires $\mathrm{exp}(O(\sqrt{\log n}))$ traces when the deletion probability is less than $0.5$~\cite{peres2017average}, building on a similar approach that uses $\mathrm{poly}(n)$ traces~\cite{HolensteinMPW08}. Subsequent work extends this idea with a more sophisticated alignment method and many technical developments, leading to the best known algorithm for average-case trace reconstruction that achieves a trace complexity of $\widetilde T_q(n) = \mathrm{exp}(O(\log^{1/3}n))$ for any deletion probability $q$ bounded away from one~\cite{holden2018subpolynomial}. Recently, an algorithm has also been proposed that achieves a polynomial number of traces in a {\em smoothed-analysis} setting that interpolates between the worst-case and average-case reconstruction problems; more specifically, in this model, a worst-case seed string is first randomly perturbed, where each bit is flipped with some probability {\em less than} $1/2$, and then the traces are all generated from this randomized string~\cite{chen2020polynomialtime}.

\subsubsection*{\textbf{Coded trace reconstruction}}
The next variation assumes that the seed string $s$ is chosen from a predefined set of possible strings (e.g., these may be \textit{codewords} from a suitable code, where it is desirable for these codewords to have an efficient construction procedure as well). For example, in DNA data storage, there is flexibility to encode the seed strings. The definition of success probability can either be the minimum over all predefined seed strings (worst-case) or the expectation over a uniformly random predefined seed string (average-case). 
\begin{quote}
\vspace{.1in}
\textbf{Coded Trace Reconstruction Problem for the Deletion Channel}\\
\textbf{Input:} A random trace-set $C$ of size $T$ generated from a seed string $s$ of length $n$ according to the deletion channel model with deletion probability $q$, where $s$ is guaranteed to be from a predefined set of possible strings, as well as the \textit{ReconstructionRate}.\\
\textbf{Output:} A string $\hat s$ such that $\hat s = s$ with success probability at least \textit{ReconstructionRate}.
\vspace{.1in}
\end{quote}

Compared to reconstructing worst-case strings, better trace complexity upper bounds are known. The improvement depends on the number of possible encoded strings, i.e., the rate of the code~\cite{abroshan2019coding, brakensiek2019coded, CGMR}.
We mention a few results that exemplify different regimes. For this discussion, we consider worst-case reconstruction, where the success probability guarantee holds for all predefined strings. It will also be convenient to frame the encoding process as adding redundancy to an arbitrary seed string. The code maps the unknown seed string $s$ of length $n$ to a new string~$s'$ of larger length $n' > n$. Applying this mapping to all possible strings generates the predefined seed strings in the coded trace reconstruction problem. The objective is to simultaneously minimize $n'$ while developing an efficient reconstruction algorithm with small trace complexity. 

We say the code has {\em redundancy} $n'-n$ equal to the number of extra characters in the encoding.  When the redundancy is small, such as $O(n/\log n)$, algorithms are known with trace complexity $\mathrm{polylog}(n)$, which is sublinear in seed string length~\cite{CGMR}. The high-level strategy is to create the new string $s'$ by concatenating many codewords. The added redundancy comes from padding the codewords with a run of zeros followed by a run of ones. For example, the codewords could have length $\Theta(\log^2 n)$ and runs have length $\Theta(\log n)$. This implies that none of the padded portions are deleted in a trace with high probability. The padding enables the algorithm to align the codeword portions in each trace. The redundancy for such a scheme is $O(n / \log n)$. After identifying the padded and codeword portions, the encoded seed string $s'$ can be reconstructed from $\mathrm{polylog}(n)$ traces.

In the larger redundancy regime, such as redundancy~$\varepsilon n$ with $\varepsilon \in (0,1)$ being a constant, an improved trace complexity of $\mathrm{exp}(O(\log^{1/3}(1/\varepsilon)))$ is achievable~\cite{brakensiek2019coded}. 
Recent work also more thoroughly studies coded trace reconstruction in the insertion/deletion channel when there are a constant number of errors or a constant number of traces~\cite{abroshan2019coding,chrisnata2020optimal,haeupler2014repeated,kiah2020coding,sabary2020error}. Before integrating these results into a DNA data storage system, certain ulterior constraints should be addressed as well. The synthesis process imposes limitations on the seed string length, and hence, the redundancy must be relatively small~\cite{ceze2019molecular, organick2018random}. Trace reconstruction is also only one part of the pipeline. The encoding and decoding schemes may need to satisfy other properties, such as error-correction capabilities~\cite{organick2018random} and enough separation between seed strings to enable clustering~\cite{rashtchian2017clustering}.

\subsubsection*{\textbf{Non-uniform error rate}}
The deletion channel model assumes that the deletion probability $q$ is fixed for all characters in the seed string. A biologically relevant modification considers varying deletion probabilities, where the position or value of each character may affect the error probability~\cite{HartungHP18}. For certain assumptions on the deletion probabilities, the current best algorithm is the same as for worst-case strings with constant deletion probability (i.e., a mean-based algorithm), and the trace complexity is asymptotically the same $\mathrm{exp}(O(n^{1/3}))$ as well. It is an important open question to extend current theoretical results to more realistic error models.

\subsection{Reconstruction of multiple seed strings}
\label{sec_dna_storage_multiple_seeds}
In many applications, the goal is to reconstruct a set of unknown seed strings (rather than a single seed string) given a set of their traces. For example, in DNA data storage, the original set of short seed strings is stored together as an unordered collection in a tube. Recovering the data results in a set of traces arising from these seed strings and involves accurately determining a large fraction of the seed strings.  Storing and retrieving a set of strings leads to interesting coding-theoretic problems as well~\cite{heckel2019characterization,heckel2017fundamental, gabrys2020mass, lenz2019anchor, lenz2018coding, lenz2019upper,  pattabiraman2020coding}.

Trace reconstruction for multiple strings has been explored recently~\cite{Ban, BanAverage,narayanan2020population}. Historically, this originates in the area of {\em population recovery}, determining an unknown distribution over a set of strings~\cite{moitra2013polynomial, wigderson2012population}. In the language of trace reconstruction, the population recovery model can be described as follows. There is an unknown set $S$ of seed strings, where only the number of strings in $S$ is given as an input. The traces are generated using a two-step process. First, a string~$s$ is chosen randomly from the set of seed strings $S$ based on the uniform distribution over $S$. Then, a trace is produced from~$s$. This process repeats $T$ times, leading to a trace-set $C$. The goal is to reconstruct at least a $1-\delta$ fraction of the strings in $S$ for a given accuracy parameter $0 < \delta < 1$. In other words, the algorithm outputs a candidate set $\widehat S$ with $|\widehat S| = |S|$, and we require that $|\widehat S \cap S| \geq (1-\delta)|S|$. The success probability is defined as the probability that $|\widehat S \cap S| \geq (1-\delta)|S|$, calculated over the random trace-set. 

Analogous to the single string problems, there are variations depending on whether a set of seed strings is an arbitrary (worst-case) or random (average-case) set of strings~\cite{Ban, BanAverage, narayanan2020population}. For the worst-case version, we define the success probability over the randomness in the trace-set generation. For the average-case version, we also include the probability of choosing random set $S$ of length $n$ strings where $|S|$ is fixed.
We remark that prior work actually considers a more intricate population recovery model for a non-uniform distribution over~$S$~\cite{Ban, BanAverage, moitra2013polynomial, wigderson2012population}. However, we use the uniform distribution because it seems more relevant to practical applications (e.g., in DNA storage, the seed strings are chosen from $S$ with approximately equal probability).
\begin{quote}
\vspace{.1in}
\textbf{Multiple String Trace Reconstruction Problem for the Deletion Channel}\\
\textbf{Input:} A random trace-set C of size $T$ generated from a set of unknown seed strings $S$ of length $n$ according the Deletion Channel model with deletion probability $q$ and an accuracy parameter $\delta$, as well as the \textit{ReconstructionRate}.\\
\textbf{Output:} A set of strings $\widehat S$ with $|\widehat S| = |S|$ such that $|\widehat S \cap S| \geq (1-\delta)|S|$ with success probability at least \textit{ReconstructionRate}.
\vspace{.1in}
\end{quote}
The output is verifiable when the original set of strings $S$ is known. In DNA data storage, the set $S$ corresponds to the set of strings that store the data, which may be used to benchmark a reconstruction algorithm.

Average-case population recovery problem has a  straightforward reduction to the single string case, both in theory~\cite{BanAverage} and in practice~\cite{organick2018random, rashtchian2017clustering}. When the seed strings are sufficiently long, they are also far apart geometrically because they have pairwise edit distance scaling linearly with their length~\cite{GMR16, navarro2001guided, schimd2019bounds}. This ensures a clear separation between groups of traces that come from one seed string rather than another. Clustering methods can accurately partition the trace-set into subsets that are generated from each individual seed string~\cite{BanAverage, rashtchian2017clustering}. Then, algorithms for the average-case problem will succeed in exactly reconstructing most of the seed strings from the clusters. When there are $|S| = M$ seed strings, the trace complexity is $\mathrm{poly}(M) \cdot \exp(O(\log^{1/3} n))$~\cite{BanAverage}.

Reconstructing a worst-case collection of seed strings is more challenging. The first approach to do so rigorously relied on subsequence statistics, and their method uses  $\exp(n^{O(M)} \cdot \sqrt{n}))$ traces~\cite{Ban}. Subsequent work improved this bound by showing how to extend the mean-based analysis for the worst-case reconstruction of a single seed string~\cite{narayanan2020population}. The resulting algorithm uses only $\exp(O(M^3 \cdot n^{1/3}))$ traces. Notice that when $M =1$, then this matches the best known bound for a single worst-case string~\cite{de2019optimal, nazarov2017trace}.

\subsection{Reconstructing Higher-Order Structures}
\label{sec_dna_storage_trees}

\begin{figure*}[t]
\centering
\includegraphics[width = 0.66\textwidth]{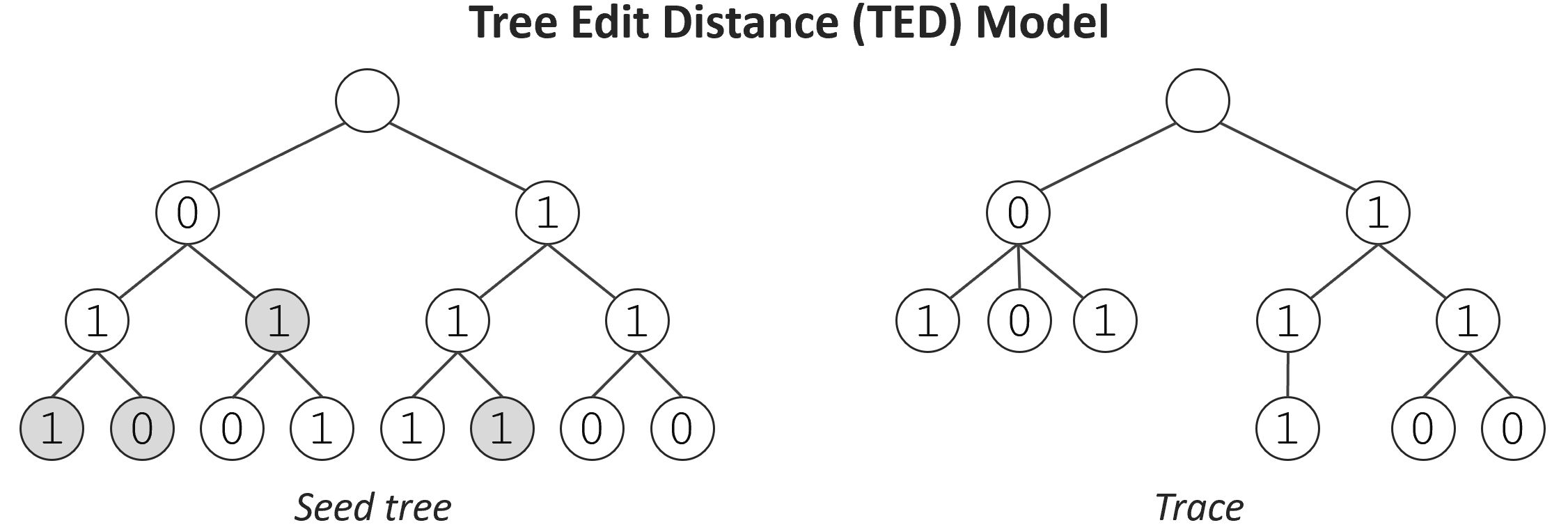}
\caption{Labeled seed tree and example trace from the Tree Edit Distance deletion channel. Gray circles indicate deleted nodes.}
\label{TED-model}
\end{figure*}

Recent work proposes a generalization of string trace reconstruction, known as {\em tree trace reconstruction}~\cite{davies2019reconstructing}. The goal is to reconstruct a node-labeled tree using traces from a channel that deletes nodes. The tree topology is known ahead of time, and learning the unknown node labels is the sole objective. They propose two deletion models that differ from each other based on how the children of a deleted node move in the tree. Figure~\ref{TED-model} depicts an example tree and trace for one of the models, which is derived from the notion of tree edit distance. When a node is deleted, its children move up to become children of the deleted node's parent. In particular, deletions still result in a connected tree. For technical reasons, the root is never deleted. The model assumes a left-to-right ordering of every level, and hence, the trees are presented in a consistent way. The tree reconstruction problem in this model generalizes string reconstruction from the deletion channel, coinciding when the tree is a path.

The tree reconstruction problem provides a vantage point to study the complexity of reconstructing higher-order structures. Perhaps surprisingly, for many classes of trees, such as complete $k$-ary trees and multi-arm stars (a.k.a. {\em spider trees}), a polynomial number of traces suffice for worst-case reconstruction~\cite{davies2019reconstructing}. This is in contrast to the string case, where the current algorithms use exponentially many traces~\cite{de2019optimal, nazarov2017trace}.  The algorithms for reconstructing complete $k$-ary trees also differ significantly from the known methods for string reconstruction. As there is more structure in the tree, combinatorial methods can be used to identify the location of certain subtrees. The algorithms make heavy use of traces that contain a root-to-leaf path of the same length as the depth of the seed tree.  If the deletion probability is constant, and the tree has depth $O(\log n)$, then such a path survives with inverse-polynomial probability. Under certain conditions, the nodes in such paths suffice to recover the corresponding labels. The algorithm for reconstructing spider trees proceeds via a mean-based approach (analogous to the worst-case reconstruction results~\cite{de2019optimal,nazarov2017trace, HartungHP18}). This involves generalizing the complex-analytic techniques to capture mean-based statistics for spider trees. It also is known that paths (a.k.a. strings) are the most difficult tree because any tree can be reconstructed using a string reconstruction algorithm with the same asymptotic trace complexity. Related endeavors study reconstructing matrices from a channel that deletes rows and columns~\cite{KMMP19} or circular seed strings from a channel that applies a random circular permutation before deleting characters~\cite{narayanan2020circular}.

Biological motivation for the tree trace reconstruction problem can be loosely attributed to the goal of identifying certain molecules that inherently have a tree-like structure. For example, recent advances have shown that tree-structured DNA is useful for bio-sensing applications~\cite{he2019fast, KT18} and storing digital information~\cite{chen2019digital}. In these applications, a variety of tree topologies have been studied. The DNA molecule could take a star-shaped form, with multiple {\em arms} connected to a shared center. The arms may be single- or double-stranded DNA, and each arm of the star contains roughly 50--100 nucleotides. Such nanostructures have been developed in the context of DNA-based nanomaterials~\cite{seeman2010nanomaterials}, using building blocks such as a 4-arm star, known as a {\em Holliday junction}~\cite{lilley2000structures}. 

The tree trace reconstruction problem arises when sequencing such tree-structured DNA. More specifically, a potential objective could be to efficiently verify that a constructed molecule has the intended shape. Nanopore devices may be able to sequence tree-structured DNA directly, providing reads that resemble traces in the tree reconstruction model. Promising initial results have been obtained for sequencing Y-shaped and T-shaped DNA~\cite{KT18}, as well as extensions to stars with up to twelve arms and certain DNA hairpin structures~\cite{chen2019digital, he2019fast}.

\subsection{Practical Trace Reconstruction Solutions}
\label{sec_dna_storage_practical}

Many theoretical trace reconstruction algorithms assume that the number of traces and the length of the seed string are both unrealistically large. Coded trace reconstruction is an exception, where simple algorithms are known with sublinear trace complexity and running time. It is possible that these theoretical algorithms can be used in practical DNA data storage systems to improve the trace complexity. A remaining challenge is combining the codes for trace reconstruction with the codes for error-correction, which is an interesting avenue for future work.

Adapting the current best theoretical algorithms for worst-case or average-case reconstruction into practical solutions seems unlikely. Instead, a promising direction is to use alignment-based methods, such as bitwise-majority alignment~\cite{BatuKannan04-RandomCase}. These perform well for the average-case problem when the deletion probability is small, and they can be efficiently implemented in near linear time. 
In one DNA data storage system, this has been successfully used when combined with certain undisclosed heuristics~\cite{organick2018random}. The idea is to start with a pointer at the beginning of each trace and move left-to-right. At each position, a majority vote is taken to determine the most likely symbol in that position. This majority symbol will be the output value of that position. Then, the pointers must be updated. If a trace agrees with the majority, then its pointer is advanced to the right by one. For the disagreeing traces, other methods must be used to guess whether the error was due to an insertion, substitution, or deletion. It is often beneficial to look ahead to the next few positions to help guess the type of error (e.g., if the next bit agrees with the majority, then the error is more likely to have been a substitution than a deletion). Depending on the type of error, the pointers for the disagreeing traces are moved appropriately.

A related approach uses a multi-sequence alignment method~\cite{edgar2004muscle} in conjunction with majority voting and certain preprocessing steps~\cite{yazdi2017portable}. Especially error-ridden traces are discarded before reconstruction. This can be based on simple criteria, such as the length of the trace or the correctness of the address portion (e.g., if the DNA primer is intact). More sophisticated methods may be used depending on the error-correcting code (e.g., parity or cyclic redundancy checks). Discarding many traces incurs a higher cost of sequencing and reconstruction, and therefore, it would be better to selectively use certain traces at different steps of the reconstruction process. 
The desired {\em ReconstructionRate} depends on the redundancy in the error-correcting codes~\cite{yazdi2017portable, organick2018random}.

Recent works take a different approach and develop ways to approximate the maximum likelihood solution~\cite{sabary2020error, srinivasavaradhan2018maximum,srinivasavaradhan2019symbolwise, srinivasavaradhan2020algorithms}. The focus here has been on developing algorithms that approximately reconstruct the seed string when given a small budget on the number of traces (e.g., 2--10). In some cases, these techniques outperform  statistical and alignment-based approaches. While this progress is promising, it is still largely an open problem to design efficient algorithms that achieve a high {\em ReconstructionRate} with a small number of traces. 

\section{Conclusion}
\label{sec_conclusion}
In this review paper, we discussed applications of the Trace Reconstruction Problem in immunogenomics and DNA data storage. We introduced new trace generation models, presented a variety of open questions, surveyed existing solutions, and discussed their applicability and shortcomings.  Given that computational immunogenomics and DNA data storage are young and rapidly expanding research areas, we expect more theoretical techniques, algorithms, and publicly available datasets to emerge in the next several years. 

\newtext{

We close with a  summary of some key open questions along with general perspectives.

\begin{itemize}
\item {\bf Maximum Likelihood vs. Trace Complexity:} Sections~\ref{sec_immunology_1} and~\ref{sec_theoretical_results} address different objectives. What are the key similarities and differences between the maximum likelihood solution and the maximum success probability solution? When does a budget on the number of traces radically influence the best reconstruction algorithm? Is there a gap between the trace complexity for computationally efficient vs. information-theoretic reconstruction?
\item {\bf Immunogenomics Models:} Throughout Section~\ref{sec_immunology_1} we have introduced several trace generation models that vary in terms of their complexity and realism. Given that these models have yet to be seriously studied, many open questions remain. Can we design polynomial-time algorithms for computing the maximum likelihood solution? Can we derive tight bounds on the trace complexity for information-theoretic reconstruction? 
\item {\bf Practical Implementations:} It remains to be seen whether an improved theoretical understanding of trace reconstruction algorithms will lead to effective empirical solutions. In Section~\ref{sec_dna_storage_practical}, we have briefly addressed some of the known practical algorithms for the deletion channel. 
What are the best performing methods in practice, in terms of trace complexity, success probability, running time, and generality?  If we want to experimentally test various algorithms, what are the important properties of benchmark datasets?
\item {\bf Confidence Measures:} Another desirable property for both immunogenomics and DNA data storage applications would be to output a measure of confidence in the reconstructed string. Is it the case that most seed strings are easy to reconstruct in practice, while only a small set of strings and traces are challenging?
\item {\bf Data Driven Models:} The models that we have surveyed involve various parameters that determine the error rate in the trace generation process. Can we experimentally determine these parameter values? Is it possible to optimize the reconstruction algorithm for the most prevalent error rates and the most realistic models? 
\item {\bf Approximate Reconstruction:} The formulation of success probability in Sections~\ref{sec_algo_formulations} and~\ref{sec_dna_storage} hinge on the requirement that the seed string is exactly reconstructed. Can we design algorithms that use fewer traces and output a candidate string within a small edit or Hamming distance of the seed string? If these additional errors can be handled with error-correcting codes, then how do approximate reconstruction algorithms compare to other approaches for coded trace reconstruction?
\item {\bf End-to-end Solutions.} Production-level DNA data  storage systems will involve a co-design of the core pipeline components. Can we develop an encoding scheme that enables efficient trace reconstruction and clustering, while also providing error-correcting capabilities and high storage density? 
\end{itemize}
}

\section*{Funding}
The work of VB was supported by the Qualcomm Institute at UCSD. The work of PAP was supported by the NIH 2-P41-GM103484PP grant and the NSF EAGER award 2032783. The work of CR was supported by the UCSD Data Science Fellowship 2019--2020. The work of YS was supported by the AAI Intersect Fellowship 2019 and the NSF EAGER award 2032783. 

\section*{Acknowledgments}

We thank the anonymous reviewers for helpful comments on earlier versions of this manuscript. We thank Sami Davis, Rex Lei, Mikl\'{o}s Z. R\'{a}cz, Jo\~{a}o Ribeiro, Omer Sabery, and Eitan Yaakobi for helpful discussions.

\ifCLASSOPTIONcaptionsoff
  \newpage
\fi



%

\bibliographystyle{IEEEtran}
\bibliography{refs}

\end{document}